\documentclass[reprint,amsmath,amssymb,aps,superscriptaddress,longbibliography]{revtex4-2}
\usepackage[dvipsnames]{xcolor}
\usepackage{graphicx}% Include figure files
\usepackage{dcolumn}% Align table columns on decimal point
\usepackage{bm}% bold math
\usepackage{hyperref}% add hypertext capabilities
\usepackage{braket}
\usepackage{float}
\usepackage{comment}
\hypersetup{%
   colorlinks = true,
   linkcolor=blue,
   citecolor=blue,
   urlcolor=blue,
}
\usepackage[T1]{fontenc}
\usepackage[utf8]{inputenc}
\usepackage{float}

\usepackage{soul}      
\usepackage[normalem]{ulem} % strikethrough text
\usepackage{units}
\usepackage{amsmath,amssymb}
\usepackage{physics}
\usepackage{multirow}
\usepackage{graphicx}
\usepackage{comment}
\usepackage[dvipsnames]{xcolor}
\usepackage{lineno}
\newcommand{\ee}             {\text{e}}
\newcommand{\Hz}             {\text{Hz}}
\newcommand{\secs}           {\text{s}}
\newcommand{\sx}             {s}

\newcommand{\micron}         {\text{µm}}

\newcommand{\thold}          {t_{\text{hold}}}

\begin{document}

\preprint{APS/123-QED}

\title{Scaling up spin complexity in ultracold assembled molecules:\\association and quantum chemistry of Cr-bearing diatomics}

\author{B. Restivo}
\affiliation{Dipartimento di Fisica e Astronomia, Universit\`{a} di Firenze, 50019 Sesto Fiorentino, Italy}
\affiliation{\mbox{European Laboratory for Non-Linear Spectroscopy (LENS), Universit\`{a} di Firenze, 50019 Sesto Fiorentino, Italy}}

\author{S. Finelli}
\affiliation{Istituto Nazionale di Ricerca Metrologica (INRiM), 50019 Sesto Fiorentino, Italy}
\affiliation{\mbox{European Laboratory for Non-Linear Spectroscopy (LENS), Universit\`{a} di Firenze, 50019 Sesto Fiorentino, Italy}}
\affiliation{Istituto Nazionale di Ottica del Consiglio Nazionale delle Ricerche (CNR-INO), 50019 Sesto Fiorentino, Italy}

\author{A. Simoni}
\email{andrea.simoni@univ-rennes.fr}
\affiliation{Univ Rennes, CNRS, IPR (Institut de Physique de Rennes)-UMR 6251, F-35000 Rennes, France}

\author{A.~Trenkwalder}
\affiliation{Istituto Nazionale di Ottica del Consiglio Nazionale delle Ricerche (CNR-INO), 50019 Sesto Fiorentino, Italy}
\affiliation{\mbox{European Laboratory for Non-Linear Spectroscopy (LENS), Universit\`{a} di Firenze, 50019 Sesto Fiorentino, Italy}}

\author{M. Zaccanti}
\affiliation{Istituto Nazionale di Ottica del Consiglio Nazionale delle Ricerche (CNR-INO), 50019 Sesto Fiorentino, Italy}
\affiliation{\mbox{European Laboratory for Non-Linear Spectroscopy (LENS), Universit\`{a} di Firenze, 50019 Sesto Fiorentino, Italy}}

\author{A. Ciamei}
\email{ciamei@lens.unifi.it}
\affiliation{Istituto Nazionale di Ricerca Metrologica (INRiM), 50019 Sesto Fiorentino, Italy}
\affiliation{\mbox{European Laboratory for Non-Linear Spectroscopy (LENS), Universit\`{a} di Firenze, 50019 Sesto Fiorentino, Italy}}

\begin{abstract}
The availability of ever more atomic elements in the ultracold regime potentially enables the extension of molecule assembly well beyond bi-alkali systems to targeted chemical species for novel quantum science and technologies. The concurrent need for single-quantum-state control and molecular constants from \emph{ab initio} calculations, especially for previously unexplored species, creates a trade-off between complexity and feasibility. Here, we address this challenge and exploit inter-isotope Feshbach spectroscopy as a tool to study three chromium-bearing diatomics ($^6$Li$^X$Cr, $^{50,52}$Cr$_2$, and $^{53}$Cr$_2$) at different levels of complexity. The extracted molecular properties, including the number of vibrational levels and the dispersion coefficients, allow us to assess quantum chemistry methods.
Moreover, our understanding of atomic scattering allows us to assemble LiCr and Cr$_2$ molecules and clarify the observed association pathways with quantum state selectivity.
Our results suggest unique prospects for the Li-Cr and Cr-Cr systems, and, more importantly, the experimental viability of a wider class of assembled high-spin molecules for many-body physics and precision measurements.
\end{abstract}

\pacs{Valid PACS appear here}
\maketitle

\section{Introduction}
\label{Sec:Intro}

Over the last two decades, molecule assembly from ultracold atoms has allowed the production of ground-state diatomics at unprecedented high phase-space densities \cite{Ni2008, Danzl2008} and even, more recently, the achievement of quantum degeneracy \cite{Valtolina2020,Schindewolf2022,Bigagli2023,shi_boseeinstein_2026}. The foreseen applications of ultracold molecular gases range from quantum computation and simulation, to ultracold chemistry and precision measurements for fundamental physics \cite{Kaufman2021,Karman2024,Cornish2024,holland,DeMille2024,PhysRevLett.132.190001}. The experimental strategy relies on a two-step process, whereby atom pairs are first associated into weakly-bound molecules across a magnetic (or optical) Feshbach resonance \cite{RevModPhys.78.1311,RevModPhys.82.1225} and later transferred to the desired deeply-bound state \cite{RevModPhys.89.015006}. While most experimental efforts have targeted bi-alkali diatomics, many other elements can nowadays be brought to ultralow temperatures \cite{schreck_laser_2021}, such as alkaline-earth metals (with only Sr$_2$ having been realized \cite{PhysRevX.13.011047}), coinage metals \cite{uhlenberg_magneto-optical_2000}, highly-magnetic transition metals and lanthanides \cite{chomaz_dipolar_2022}, with cooling methods being extended to species of increasing complexity~\cite{PhysRevA.89.041401,PhysRevLett.129.223401,prudnikov_new_2016,yu_magneto-optical_2022,dzuba_time_2021,eustice_magneto-optical_2025,paladugu_chirped_2025,padilla-castillo_large_2025,moller_magneto-optical_2025,mieloch_exploring_2026,zheng_simulated_2026}. 
Assembly techniques can in principle be extended to ever more complex molecular classes targeting specific properties and isotopologues, including radioactive ones \cite{klos_prospects_2022,polet_pt-odd_2024,elhalawani_what_2026,ciamei_ultracold_2026,RbSrFFR,ladjimi_diatomic_2024,tiberi_searching_2024,vayninger_magneto-optical_2025,marc_candidate_2023}, some of which are under active investigation. 
However, due to the lack of experimental data of many species of interest, the mere existence of association channels does not guarantee access to such molecules, as single-quantum-state control at the association stage, as well as accurate theoretical predictions of spectroscopic constants at the transfer stage, are required. The former demands quantum number assignment, and the latter state-of-the-art quantum chemistry. Thus, a trade-off between complexity and feasibility is expected.

\begin{figure*}[ht!]
\centering
  \includegraphics[width=\textwidth]
  {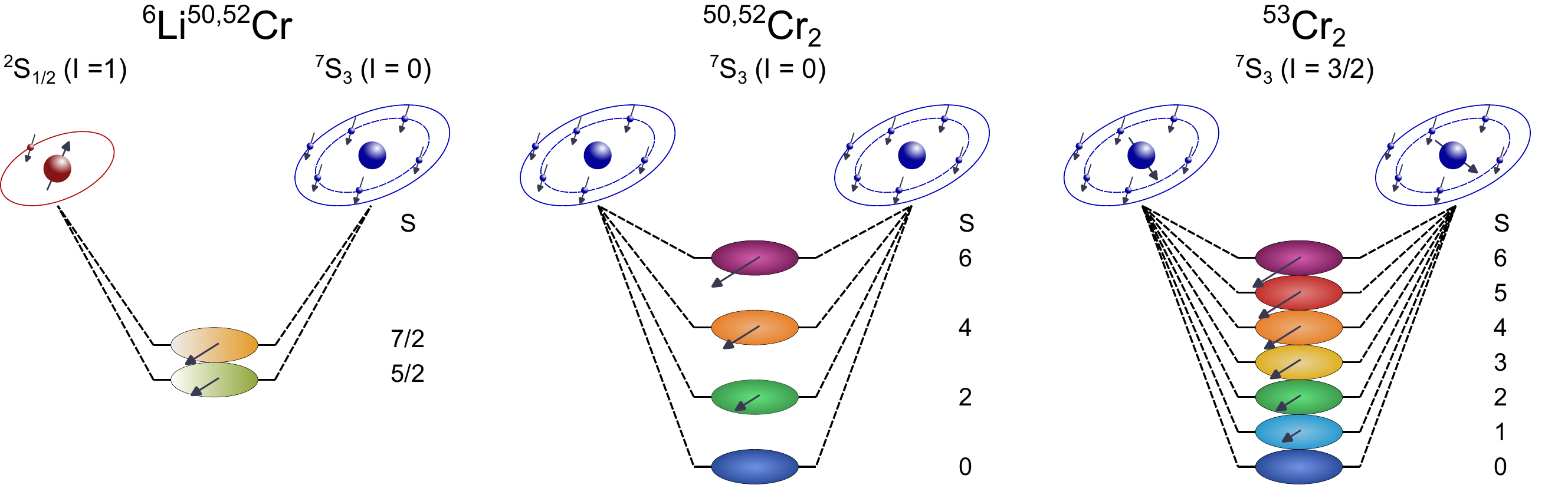}
  \caption{Sketch of the molecular states of electron spin $S$ accessible from $s$-wave scattering of the different atomic combinations here investigated. $^6\text{Li}{}{^X\text{Cr}}$ has 2 high-spin multiplicities both observable under $s$-wave scattering \cite{PhysRevLett.129.093402, PRXQuantum.5.020358}. The chromium dimer features 7 spin states, all of which are experimentally accessible from distinguishable $^{53}$Cr fermions, while only the even-$S$ states are addressed with identical $^{50,52}$Cr bosons. We introduce the color scale for different $S$ used throughout the paper. Shading of the color represents polarized charge densities.} 
  \label{fgr:SketchMols}
\end{figure*}
Transition-metal chromium, featuring high spin but zero orbital angular momentum, has been recently identified as a prime candidate to assemble open-shell dimers with non-zero electron spin, unattainable in the singlet ground-state of bi-alkalis.
Successful formation of LiCr molecules has recently paved the way to production of polar paramagnetic molecules for quantum simulation \cite{PRXQuantum.5.020358}, and YbCr, where the combination of high-spin Cr and relativistic Yb grants access to $\Omega$-doublets, has been proposed for next-generation charge-parity violation searches \cite{ciamei_ultracold_2026}.
Furthermore, the antiferromagnetic ground-state of Cr$_2$, due to its multiple metal bonding, has been experimentally investigated \cite{Cr1983,casey_negative_2002} and represents a workhorse in the quantum chemistry community~\cite{larsson_chromium_2022,Olivares-Pilon:2024azn}. The excited paramagnetic spin states of interacting ground-state Cr are computationally challenging for \emph{ab initio} methods \cite{desmarais_magnetic_2000,PhysRevA.69.030701} and have never been observed. Experimental information, even at the dimer level, can provide insight into chemical bonding in the presence of multiple spin-exchange terms \cite{buchachenko_electronic_2010,patkowski_first-order_2018} and cluster formation \cite{morse_clusters_2002,wurdemann_density_2015}.

In this work, we assess the viability of different Cr-bearing diatomics beyond the bi-alkali paradigm at three different levels of complexity: heteronuclear $^{6}$Li$^{X}$Cr isotopologues (with $X = 50, 52, 53$), $^{50,52}$Cr$_2$ homonuclear dimers and finally $^{53}$Cr$_2$, see Fig.~\ref{fgr:SketchMols}. In the first case, similar to bi-alkalis, we deal with only two electron-spin multiplicities but with higher angular momenta ($S~=~5/2,7/2$) \cite{PRXQuantum.5.020358}. In the second case we study bosonic Cr isotopes, featuring 7 spin multiplicities ($S=0,...,6$), where due to exchange symmetry we are only sensitive to even-$S$ states. In the latter case we exploit fermionic $^{53}$Cr to introduce high nuclear spin ($I=3/2$) and sensitivity to all spin states. We employ inter-isotope Feshbach spectroscopy and fitted mass-scaled collisional models to determine the number of vibrational levels supported by electronic states of different spin-multiplicities. We use these results (together with dispersion coefficients) to assess quantum chemistry methods reported in literature. In parallel, despite the increased spin complexity, our understanding of atomic scattering properties allows us to produce LiCr and Cr$_2$ diatomics with quantum state and isotopologue selectivity.

The paper is organized as follows. In Sec.~\ref{Sec:Exp}, we describe the experimental procedures used for atomic sample preparation and Feshbach spectroscopy. In Sec.~\ref{Sec:LiCr} we develop a full mass-scaled model for Li-Cr and benchmark available \emph{ab initio} potentials for ground sextet and excited octet states. In Sec.~\ref{Sec:50Cr50Cr}, we report the experimental results of the Feshbach spectroscopy on identical $^{50}$Cr bosons. Combined with $^{52}$Cr data from previous work \cite{PhysRevLett.94.183201} and herein refined, this allows us to derive a mass-scaled collisional model with multiple spin-channels and to test quantum chemistry methods. In Sec.~\ref{Sec:53Cr53Cr}, we present our results for the more complex $^{53}$Cr via Feshbach spectroscopy on distinguishable fermions, together with our unambiguous assignment and the inferred scattering lengths.
In Sec.~\ref{Sec:Prospects}, we discuss the formation of Cr-bearing molecules as well as the implications on many-body physics in Li-Cr and Cr quantum gases. 
Finally, in Sec.~\ref{Sec:Conclusions}, we summarize our conclusions and provide an outlook.

\section{Experimental strategy}
\label{Sec:Exp}

Our experimental apparatus and the strategies employed to produce $^6$Li-$^{53}$Cr Fermi mixtures have been already reported in our previous works \cite{PhysRevA.101.063602, PhysRevA.106.053318, PhysRevLett.129.093402,PRXQuantum.5.020358}. A recent upgrade of the chromium laser setup, that will be described elsewhere \cite{LiCrMixtures}, allows us now to realize, through an optimized all-optical procedure, ultracold samples of up to $2.5 \times 10^5$ atoms of the bosonic isotopes $^{52}$Cr and $^{50}$Cr.
Notably, bosonic $^{50,52,54}$Cr isotopes lack nuclear spin $(I=0)$, while it is present in the more complex fermionic $^{53}$Cr $(I = 3/2)$. 
Given the absence of a hyperfine structure in bosons, the cooling transition $^7S_3 \rightarrow {}^7P_4$ is closed except for a small leak to the metastable $^5D_{3,4}$ states which are repumped via the $^7P_3$ state into the cooling cycle. Instead, for $^{53}$Cr, the cooling transition $^7S_3 (F=9/2) \rightarrow {}^7P_4 (F'=11/2)$ needs to be closed by additional repumping the excited hyperfine levels ($F=7/2,5/2,3/2$) of the $^7S_3$ ground state~\cite{PhysRevA.101.063602, PhysRevA.106.053318}. Out of the three bosonic isotopes, $^{54}$Cr is not addressed in the present work because of the accidental coincidence of its cooling transition with the much more abundant $^{52}$Cr one \cite{Furmann2005}.
Upon adjusting the final part of the evaporation stage, optically-trapped chromium gases can either be produced alone or in combination with the $^6$Li component, the latter comprising up to $6\times10^5$ atoms.

Here we exploit these new atomic mixtures to experimentally extract information about the molecular systems as well as to infer atomic scattering properties of both heteronuclear $^6$Li${}^X$Cr and homonuclear $^X$Cr$_2$ isotopologues $(X=50,52,53)$ by means of inter-isotope Feshbach spectroscopy \cite{RevModPhys.82.1225}.
In particular, in the $^6$Li$^X$Cr case we focus on the $^6$Li$\ket{\text{a}}$-$^X$Cr$\ket{\text{a}}$ combinations, where the atomic states are ordered with increasing Zeeman energy. For bosonic $^X$Cr$_2$ we also explore Feshbach resonances (FRs) in the $^X$Cr$\ket{\text{a}}$-$^X$Cr$\ket{\text{a}}$ scattering channel, whereas for fermionic chromium we employ $^{53}$Cr$\ket{\text{a}}$-$^{53}$Cr$\ket{\text{b}}$ spin mixtures. The desired system is produced by adjusting the final stage of the evaporation, combined with a sequence of state- and species-selective optical blasts and radio-frequency pulses that remove undesired spin states or atomic components.

\begin{figure}[t!]
\centering
\includegraphics[width=\columnwidth]{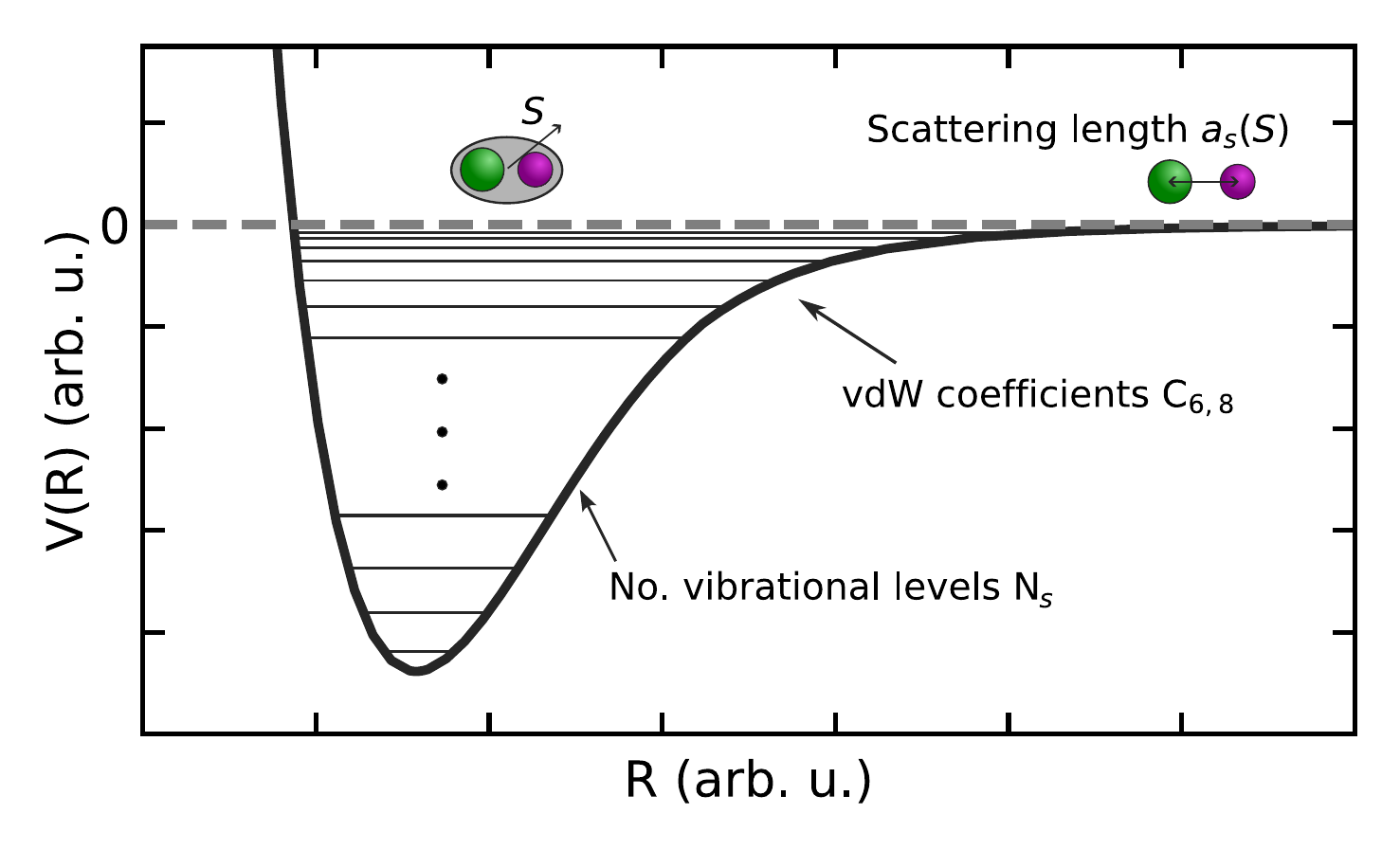}
\caption{Sketch of a molecular potential of electron spin $S$ together with the associated information extracted in this work: molecular properties (number of vibrational levels $N_{S}$, van der Waals coefficients $C_{6,8}$) and atom-atom scattering properties (the $s$-wave scattering length $a_s(S)$).}
\label{fig:potential}
\end{figure}

By switching between different isotopes, and assuming the Born-Oppenheimer approximation, we effectively probe the very same electronic potentials with different reduced masses, which results in a mass shift of the weakly-bound levels inducing the FRs. This variation, depending on the relative mass change and number of vibrational levels, together with the presence or absence of a nuclear spin in Cr, yields a variation in resonance locations and pattern. In the Li-Cr system, where the reduced mass is only slightly varied, relative small shifts of the resonance locations are expected, apart from specific regions with high sensitivity to scattering parameters. On the other hand, in the Cr-Cr system, the relative mass shift of up to 5\% yields larger resonance shifts and different patterns. Given the weakly-bound nature of levels inducing these resonances, hyperfine and electron spin-spin interactions are determined by atomic constants to a good approximation, and the mass shift contribution to the isotopic shifts can be singled out to provide information on the number of vibrational bound states supported by the interaction potential. Fig.~\ref{fig:potential} shows a sketch of a molecular potential together with the molecular and atomic quantities extracted in this study.

Loss spectroscopy is performed by exploiting three independent, concentric sets of magnetic coils, oriented along the vertical direction, whose currents are stabilized through a LEM device \cite{PRXQuantum.5.020358}, and that allow us to scan the magnetic field $B$ from zero up to 1600 G. The first two main sets of coils (FESH1 and FESH2) allow to produce large bias fields with coarse $B$-field steps of about 50 mG and to generate a levitating magnetic field gradient, respectively, whereas the third set (FESH3) enables fine steps below 1 mG and a dynamic range of about 2 G. The setup is first employed for rapid, low-precision scans over wide magnetic field regions. Starting from the magnetic field value ($B_{\text{in}}\sim1300$ G)  set during the evaporation stage \cite{LiCrMixtures}, we reach the desired field by linearly ramping the current of the FESH1 coils over typically 200-250 ms. We then hold the samples for about 2 seconds in the optical dipole trap (ODT), during which we apply linear $B$-field sweeps by means of our FESH3 coils. We then ramp back to $B_{\text{in}}$, where state-selective absorption imaging is performed to determine the final atom number. The presence of FRs within the explored range is signaled, as usual, by enhanced atom losses \cite{RevModPhys.82.1225}. Once a resonance is found through this low-precision measurement, we perform a more refined scan by monitoring atom losses as a function of the FESH3 current (at constant FESH1 and FESH2 current value). The hold time, ranging from 100 ms to 4 s, is optimized to avoid saturation effects of the loss spectrum.

To increase our accuracy, the $B$-field locations of a subset of the FRs are refined by exploiting a spin-polarized Li gas as a magnetometer. To this end, we follow the same experimental sequence but preparing a pure Li$\ket{\text{a}}$ sample in the ODT, perform radio-frequency spectroscopy onto the  $\text{Li}\ket{\text{a}}\leftrightarrow\text{Li}\ket{\text{b}}$ transition at the coils currents where loss peaks are observed, and extract the corresponding magnetic fields via the Breit-Rabi formula. This results in a typical uncertainty of about 20 mG for a large subset of the observed FRs. Furthermore, we use these measurements to derive a calibration of the magnetic field produced by our coils with about 100 mG uncertainties, which is exploited for the remaining features.

Concerning the Cr-Cr system, besides guiding the initial quantum number assignment via the observed loss features, we further assess the predictability of our final model in terms of resonance strengths from accurate measurements of magnetic field width of specific FRs. Indeed, while we include the contribution of direct electron spin-spin interaction to spin-spin couplings, we choose to neglect second-order spin orbit on the theory side. On the one hand, this choice drastically simplifies our fit to the experimentally determined FRs locations, and the attainment of an unambiguous assignment under this assumption is indeed one of the main results of the present paper. On the other hand, our approach requires an independent check of FRs strengths. To this end, we select one FR for each of the two new $^{50}$Cr$_2$ and $^{53}$Cr$_2$ isotopologues herein explored and perform an accurate measurement of their magnetic-field width. This is achieved by precisely locating the zero-crossing of the scattering length through a new protocol to measure elastic scattering rate, see Appendix \ref{app:UhlOrn} for details.

\section{Mass-scaled L\MakeLowercase{i}-C\MakeLowercase{r} model and assessment of \emph{ab initio} results}
\label{Sec:LiCr}

\begin{figure*}[ht!]
\centering
  \includegraphics[width=\textwidth]
  {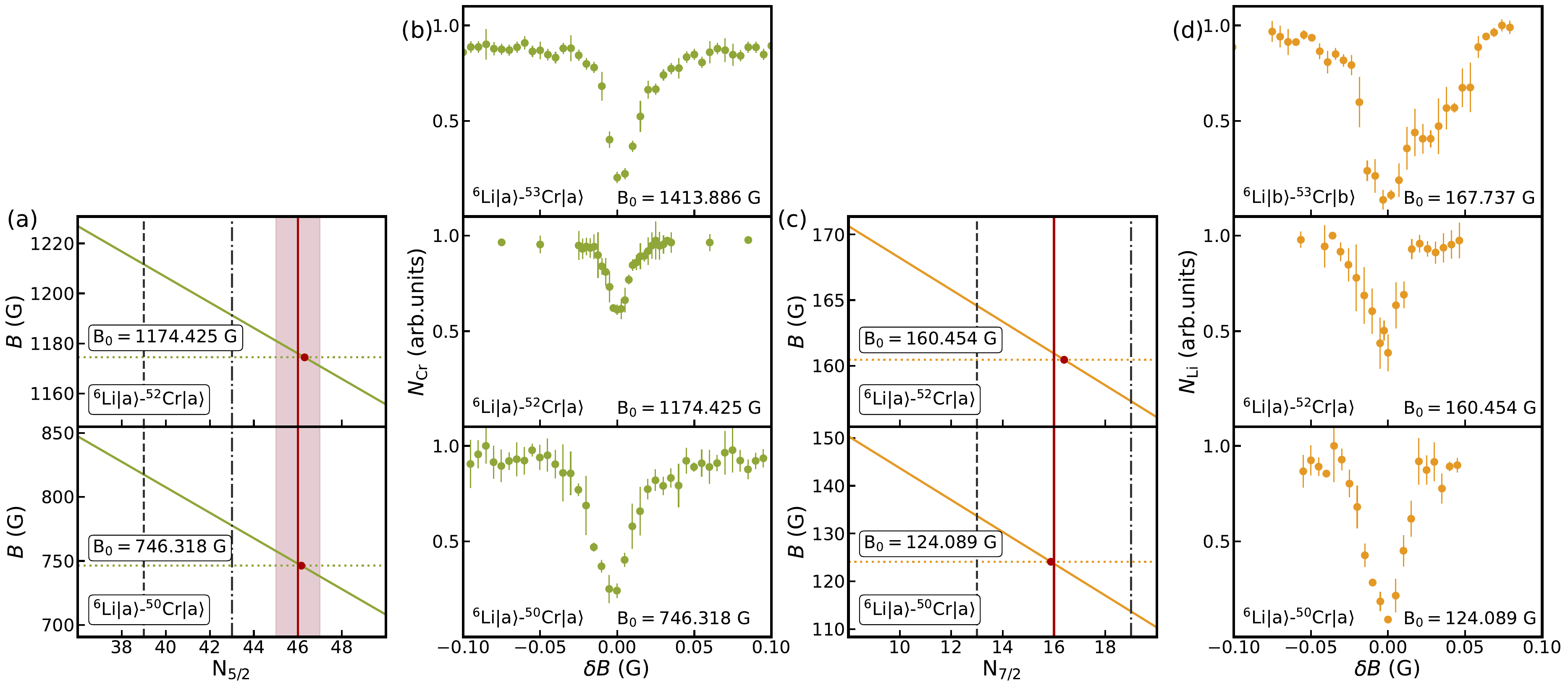}
  \caption{
  Loss features and isotopic shifts of selected FRs in Li-Cr mixtures. In panels (a) and (c) the green and orange lines represent the expected variation of the $B$-field location of the FRs as a function of the number of bound states. The red lines are the theory predictions of our Ref.~\cite{PRXQuantum.5.020358} (with the error represented by the corresponding red shadow regions), while the dashed (dotted-dashed) lines are the prediction of Ref.~\cite{Jeung_2010} (Ref.~\cite{DENG201669}). The red points correspond to the magnetic fields values of the isotope-shifted Feshbach resonances measured experimentally.
  Panels (b) and (d) show high-precision loss-spectroscopy scans as a function of the magnetic field detuning $\delta B$, for the $^6$Li-${}^X$Cr (see legends) Feshbach resonances induced by the sextet and the octet potential, respectively.
  } 
  \label{fgr:OverviewMerged}
\end{figure*}

We measure the isotopic shifts of selected Feshbach resonances in Li-Cr mixtures, induced by changing the Cr isotope, to gain information on the molecular properties for both the electronic ground state $^6\Sigma$ ($S=5/2$) potential and the first excited state $^8\Sigma$ ($S=7/2$) \cite{PRXQuantum.5.020358}, as well as to develop a complete mass-scaled quantum collisional model. In this system, where the reduced mass is only slightly varied upon changing the Cr isotope, relatively small shifts in the resonance location are expected, apart from specific regions with high sensitivity to scattering parameters. The extracted molecular properties, namely the overall number of vibrational levels and dispersion coefficients, are used to benchmark recent theoretical predictions based on different quantum chemistry methods~\cite{PRXQuantum.5.020358,DENG201669,Jeung_2010}, which are summarized in Table \ref{tab:LiCrNvib}.

\begin{table}[b!]
  \centering
  \caption{Theory predictions and experimental results of Li-Cr number of vibrational states $N_S$ and dispersion coefficients $C_{6,8}$ calculated for the $^6$Li-$^{52}$Cr combination. The dispersion coefficients are expressed in atomic units, where $E_h$ is the Hartree energy and $a_0$ is the Bohr radius.}
  \label{tab:LiCrNvib}
    \begin{ruledtabular}
      \begin{tabular}{c cc cc}
         &  \multicolumn{2}{c}{$N_S$}  & $C_6$ & $C_8$\\
         &  $^{6}\Sigma$& $^{8}\Sigma$ & ($E_h\cdot a^6_0$) & ($E_h\cdot a^8_0$)\\[1mm]
         \hline 
          \noalign{\vskip 1mm}
         MRCI \cite{Jeung_2010} & 39 & 13& [1002, 810]$^{*}$ & 1.018$\times 10^5$\\ 
         MRCI+Q \cite{DENG201669}& 44 & 19 & & \\ 
         CCSDTQ \cite{PRXQuantum.5.020358} & 46(1) & 16& 954 & \\
         This work & 46& 16 & 925.1(4) & 73.62(38)$ \times 10^3$\\[1mm]
       \end{tabular}
    \end{ruledtabular}
    \makebox[\linewidth][l]{\footnotesize $^{*}$ Calculated from approximations based on atomic polarizabilities.}
\end{table}

The $^6$Li-$^{53}$Cr Feshbach spectrum, previously measured in our group \cite{PhysRevLett.129.093402}, exhibits resonances induced by energy crossings with molecular levels of almost pure sextet or octet symmetry. Hence, we identified and targeted a single $s$-wave resonance per electron-spin channel to efficiently extract all necessary information on this system. In particular, we selected the resonance at 1413.886 G in the $^6$Li$\ket{\text{a}}$-$^{53}$Cr$\ket{\text{a}}$ channel for the sextet potential, and the resonance at 167.737 G in the $^6$Li$\ket{\mathrm{b}}$-$^{53}$Cr$\ket{\mathrm{b}}$ channel for the octet potential, which provide the highest sensitivity to mass shifts. 
In both cases, the main coupling mechanism is spin-exchange, which gives rise to the splitting between the sextet and octet potentials in the exchange region, and can induce transitions between atomic hyperfine states.

The isotope-shifted resonances we search for possess the same quantum numbers apart from the nuclear spin contribution of Cr. To improve the precision, we measure the magnetic-field shifts of both resonances using $^{50}\text{Cr}$ and $^{52}\text{Cr}$ isotopes.
To predict the magnetic field range in which the corresponding $^6$Li-$^{50,52}$Cr resonances are expected to appear, we first exploited an asymptotic bound-state model (ABSM) \cite{PhysRevA.82.042712} with the initial guess of number of vibrational levels from the most recent \emph{ab initio} results of our Ref.~\cite{PRXQuantum.5.020358}. In the model, the binding energies and the scattering lengths of the $^6$Li-$^{53}$Cr system are fixed to their known experimental values \cite{PhysRevLett.129.093402}, while the number of bound states is allowed to vary. The approximately linear relationship between the expected position of the FRs poles and the number of bound states for the $^6$Li-$^{50,52}$Cr combinations is shown in Figs.~\ref{fgr:OverviewMerged}(a) and (c).

Experimentally, we perform a magnetic field scan with spin-polarized mixtures in an interval of a few Gauss, centered around the values predicted via mass-scaling assuming $46 \pm 1$ (exactly 16) vibrational levels for the sextet (octet) potential. The scans are performed with typical atomic peak densities ranging from $0.9$ to $1.5\times10^{12}~\text{cm}^{-3}$ for $^6$Li and from $0.2$ to $1\times10^{12}~\text{cm}^{-3}$ for Cr, at temperatures between $300$ nK and $500$ nK. The resonances were successfully found, and their $B$-field location accurately measured by magnetometry on the Li$\ket{\text{a}} \leftrightarrow \ket{\text{b}}$ transition. The observed loss features are shown in Fig.~\ref{fgr:OverviewMerged} for sextet (panel b) and octet (panel d) resonances. 
The results from the ABSM model were refined by rigorous coupled-channel scattering calculations, that confirm the physical picture and the corresponding assignment (see Table~\ref{tab:FRsLiCr} of Appendix~\ref{Appendix:LiCrFRs} for further details).

Our experimental results allow us to pinpoint the exact number of vibrational levels for both spin symmetries. We obtain exactly 46 bound states for the sextet potential and 16 bound states for the octet potential for the isotopologue $^6$Li-$^{52}$Cr. In both cases we find excellent agreement with the theory predictions based on coupled cluster methods up to quadrupole excitations (CCSDTQ) \cite{PRXQuantum.5.020358}, while multi-reference configuration interaction methods (MRCI and MRCI+Q) show larger deviation from experimental data for both spin states~\cite{DENG201669, Jeung_2010}, see Table~\ref{tab:LiCrNvib} and Figs.~\ref{fgr:OverviewMerged}(a) and (c). Interestingly, the MRCI method consistently underestimates the number of vibrational levels for both spin states, while the MRCI+Q method better approaches the experimental result for the sextet state and even overestimates the octet value.

Finally, we performed an accurate inter-isotope fit of all observed resonances using coupled-channel calculations based on the potential-energy curves (PECs) of Ref.~\cite{PRXQuantum.5.020358}. By optimizing the potential well depths and dispersion coefficients, we obtained a quantitative mass-scaled model for Li-Cr. The resulting dispersion coefficients are shown in Table~\ref{tab:LiCrNvib}, while the $s$-wave scattering lengths for all isotopic combinations are reported in Table~\ref{tab:scatlengthsLiCr}. The predicted van der Waals coefficients all lie within 10\% from the fitted values, with the perturbative result from Ref.~\cite{PRXQuantum.5.020358} providing the best agreement.

\begin{table}[t!]
  \centering
  \caption{Octet and sextet $s$-wave scattering lengths of all Li-Cr isotopologues in units of Bohr radii.}
  \label{tab:scatlengthsLiCr}
    \begin{ruledtabular}
      \begin{tabular}{cccccc}
        Isotope  & $^{50}$Cr  & $^{52}$Cr & $^{53}$Cr & $^{54}$Cr & S\\ [1mm]
        \hline 
         \noalign{\vskip 1mm}
         $^{6}$Li   & 46.888(2) & 43.085(2) & 41.383(2) & 39.791(2) &  \multirow{2}{*}{$7/2$} \\ 
         $^{7}$Li   & 34.6604(8) &  30.5977(9)  & 28.6478(10) & 26.7406(12) &\\ [1mm]
         \hline
          \noalign{\vskip 1mm}
         $^{6}$Li   & 30.094(7) & 20.431(8) & 15.172(9) &  9.286(12) & \multirow{2}{*}{$5/2$}  \\ 
         $^{7}$Li   & -2.42(2) &  -48.57(7)  & -124.0(2) & -925(8) &\\ [1mm]
       \end{tabular}
    \end{ruledtabular}
    \vskip -10pt
\end{table}

\section{Testing quantum chemistry methods on even-$S$ states of C\MakeLowercase{r}$_2$}
\label{Sec:50Cr50Cr}

\begin{figure*}[ht!]
\centering
  \includegraphics[width=\textwidth]
  {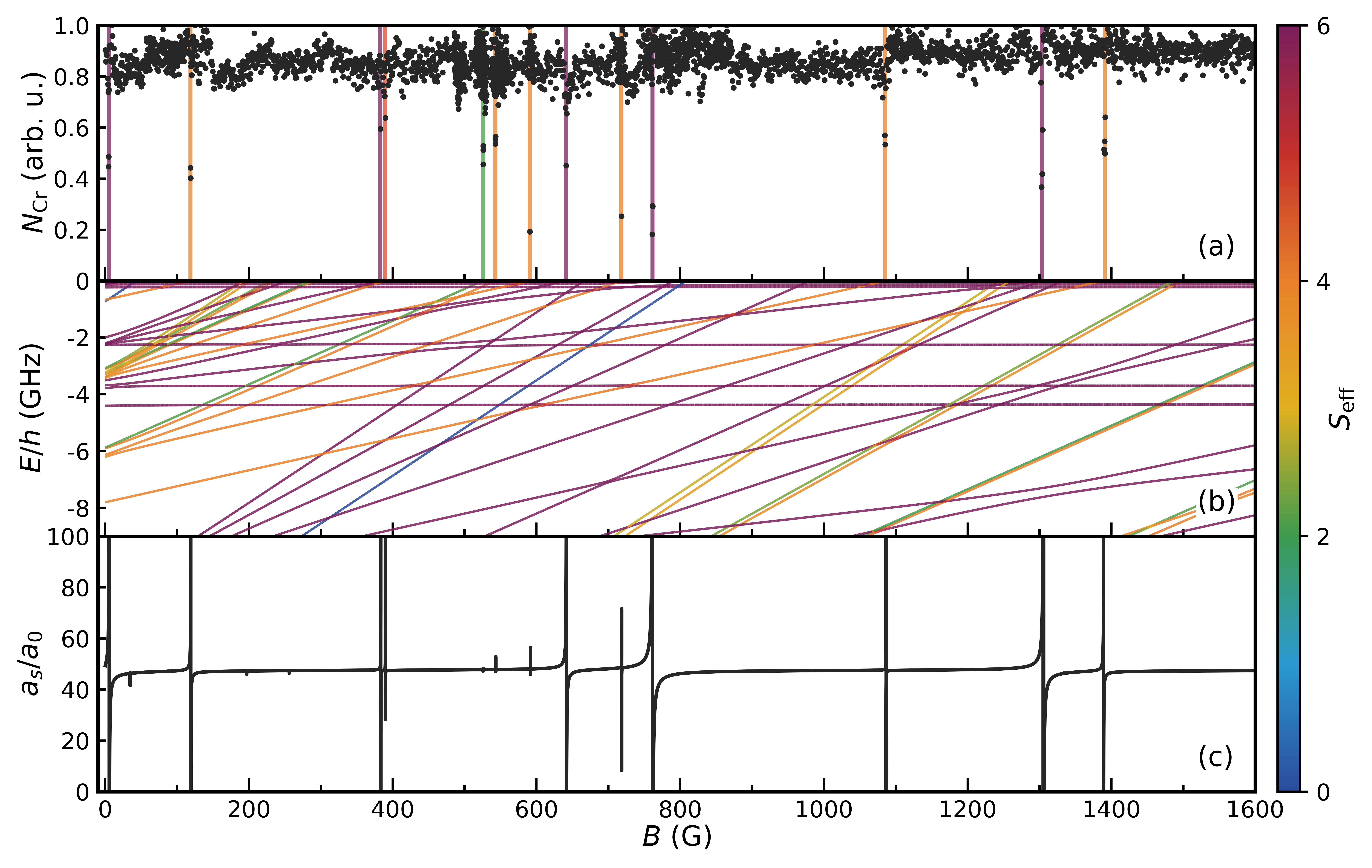}
  \caption{Comparison between the experimental and theoretical Feshbach spectra for the ${}^{50}\text{Cr}\ket{\text{a}}$-$ {}^{50}\text{Cr}\ket{\text{a}}$ system. (a) Overview of the low-precision loss spectroscopy scan. The vertical lines indicate the resonances positions, with each color corresponding to the effective spin (see the color bar on the right). (b) Energy of the molecular levels (referenced to the scattering channel energy) with the same color code as (a). The total angular momentum has a projection $M=-6$ on the quantization axis. (c) The $s$-wave scattering length $a_s/a_0$, where $a_0$ is the Bohr radius, as a function of the magnetic field $B$. Theoretical calculations include partial waves up to $\ell_{\rm max}=6$ 
  } 
  \label{fgr:50CrOverview}
\end{figure*}

We now move to the notably more complex Cr$_2$ dimer, that represents a workhorse for the quantum chemistry community \cite{larsson_chromium_2022,Olivares-Pilon:2024azn}. This is characterized by seven electronic potentials $^{2S+1}\Sigma$, corresponding to total electronic spin $S=0,\ldots,6$. In terms of molecular physics, Cr$_2$ involves the interplay between the atomic outer $s$-shell, dominating the highest spin state ($S=6$), and the inner half-filled $d$-shell, contributing to the shelved anti-ferromagnetic ground state ($S=0$), with an expected transition between the two limits across intermediate spin values. In terms of atomic collisions, high-spin, orbitally-isotropic Cr features a significant magnetic anisotropy but lacks an electric one, as opposed to highly-magnetic lanthanides such as Dy and Er \cite{PhysRevX.10.041005}. For the bosonic isotopes, the nuclear spin is zero, such that only electron spin-spin induced resonances are allowed. Moreover, in the case of identical bosons, only symmetric electronic states contribute, corresponding to even values of $S$. 

In order to test quantum chemistry methods on the chromium dimer, we first address the simpler case of bosonic isotopes. The FR spectrum of $^{52}\text{Cr}\ket{\text{a}}$-$ ^{52}\text{Cr}\ket{\text{a}}$ was already characterized by Pfau's group, and all the observed resonances were assigned \cite{PhysRevLett.94.183201}. Here, we perform Feshbach spectroscopy on the less abundant $^{50}$Cr isotope in the same $\ket{\text{a}}$-$\ket{\text{a}}$ channel, and combine it with the $^{52}$Cr results.
We first record low-precision scans of the magnetic field from 0~G up to 1600~G, see Fig.~\ref{fgr:50CrOverview}(a). The measurements are performed with $^{50}$Cr densities ranging from $0.6$ to $2.5\times10^{12}~\mathrm{cm}^{-3}$ and temperatures between $370$ and $600$ nK. To explore the low $B$-field region, we add a vertically oriented ODT at 1560 nm to increase the axial confinement beyond the one provided by the magnetic curvature of the FESH1 coils. 
Moreover, both FESH1 and FESH2 coils are used simultaneously in Helmholtz configuration to scan the highest magnetic field range of 1450-1600 G. 
The experimental results for the FRs magnetic field locations are summarized in Table \ref{tab:FRsCr50} of Appendix \ref{Appendix:CrCrFRs}, where the reported errors depend on the method employed to determine the magnetic field.
In order to minimize systematic shifts between $^{50}$Cr and $^{52}$Cr resonances, we further measured with higher accuracy at lower temperature the location of previously reported resonances from the Stuttgart group \cite{PhysRevLett.94.183201}, summarized in Table~\ref{tab:52CrComparison}  of Appendix~\ref{Appendix:CrCrFRs}.

\begin{table}[h!]
  \centering
  \caption{Number of bound levels $N_{S}$ for different spin multiplicities referenced to $^{52}$Cr$_2$ and dispersion coefficients $C_{6,8}$ expressed in atomic units.}
  \label{tab:NvC6C8CrCr}

    \begin{tabular}{lccccccc}
    \hline
    \vspace{0mm}
    \\[-2.5mm]
    &&&&$N_{S}$ &&\\
      $\qquad \qquad \qquad$ $S$ & 6 & 5 & 4 & 3 & 2 & 1 & 0 \\
      \hline
      \noalign{\vskip 1mm}
      CASSCF/ \cite{PhysRevA.69.030701}     & 30 & 54 & 65 & 75 & 88 & 100 & 112 \\
      CASPT2 &&&&&&& \\
      \noalign{\vskip 1mm}
      ROHF+  \cite{Cr2Piotr}    & 32 &    &    &    &    &     &     \\
      TDDFT \\
      \noalign{\vskip 1mm}
      CIPT2+Q/DK  \cite{Cr2Piotr}    & 35 &    &    &    &    &     &     \\
      \noalign{\vskip 1mm}
      RCCSD(T)/   \cite{Cr2Piotr}   & 35 &    &    &    &    &     &     \\
      DK+b \\
      \noalign{\vskip 1mm}
      CCSDT \cite{NstateS6CrCr} & 36(1) & & & & & & \\
      \noalign{\vskip 1mm}
      Experiment  \cite{Olivares-Pilon:2024azn} &&&&&&& 112 \\
      This work & 36 & $57^{+19}_{-20}$ & 77 & $86^{+8}_{-8}$ & 95 & $104^{+7}_{-8}$ & \\[1mm]
          \hline
    \end{tabular}

    \vspace{2mm}

    \begin{tabular}{l@{\hspace{3mm}}c@{\hspace{3mm}}c}
      & $C_6$ ($E_h\cdot a^6_0$) &  $C_8$ ($E_h\cdot a^8_0$)\\
      \hline
      \noalign{\vskip 1mm}
      Semiempirical  \cite{PhysRevA.69.030701}    & 745(55) &  \\
      \noalign{\vskip 1mm}
      TDDFT  \cite{Cr2Piotr}    & 626.0 &  $3.27 \times 10^4$\\
      \noalign{\vskip 1mm}
      CIPT2+Q/DK  \cite{Cr2Piotr}    & 845.0 &  \\
      \noalign{\vskip 1mm}
      RCCSD(T)/DK+b  \cite{Cr2Piotr}    & 780.0 & \\
      \noalign{\vskip 1mm}
      Experiment \cite{PhysRevLett.94.183201} & 733(70) & $75^{+90}_{-75} \times 10^3$ \\
      \noalign{\vskip 1mm}
      This work  & $732.35(19)$ &  $3.968(21) \times 10^4$ \\
      \hline
    \end{tabular}
\end{table}

Given the observed Feshbach spectra of the $^{50}\text{Cr}$ and $^{52}\text{Cr}$ in the $\ket{\text{a}}\text{-}\ket{\text{a}}$ channel, inferring the number of vibrational levels requires the assignment of resonances' quantum numbers. Based on the PECs reported in Ref.~\cite{PhysRevA.69.030701} for all spin channels except for $S=6$, for which we use the more recent RCCSD(T) one from Ref.~\cite{Cr2Piotr}, by employing the potential well depths and dispersion coefficients $C_6$ and $C_8$ (spin-independent) as fitting parameters, we could assign all the observed $^{50}\text{Cr}\ket{\text{a}}$-$ {^{50}\text{Cr}}\ket{\text{a}}$ resonances. Our quantum collisional model, including direct spin-spin interactions and bound molecular levels up to $\ell=6$ rotational quantum number, captures well the experimental features both in terms of magnetic-field locations and strengths. Overall, we obtain good agreement with the experimentally determined fields, typically within 1\,G error, see Table \ref{tab:FRsCr50} in Appendix \ref{Appendix:CrCrFRs}. The observed resonances are induced by the $S=$ 2, 4 and 6 potentials and are marked in Fig.~\ref{fgr:50CrOverview}(a) with green, orange and purple lines, respectively. Fig.~\ref{fgr:50CrOverview}(b) correspondingly shows the energy of the molecular levels (referenced to the scattering channel's one) with the same color code, while Fig.~\ref{fgr:50CrOverview}(c) shows the theoretically-inferred magnetic field dependence of the $s$-wave scattering length. Since the entrance channel is immune to two-body losses, the scattering length genuinely diverges at the pole of each observed resonance. Unresolved or only partially resolved resonances in the figure arise solely from the final step of the numerical calculation.

As a result of the inter-isotope fit, we are able to fix the number of bound states $N_S$ for the even-$S$ excited potentials, in addition to the already established $S=0$ ground-state \cite{Olivares-Pilon:2024azn}. To assess the possible impact of additional interactions not included in our theoretical framework, such as the second-order spin-orbit coupling, we further check the predictive power of our model also in terms of magnetic-field width, on top of resonant $B$-field. Exploiting the FR at 1390 G, we measure $\Delta_{\text{exp}}=130(6)$~mG, see Appendix \ref{app:UhlOrn}, which is in excellent agreement with the theory value $\Delta_{\text{th}}=127$~mG.

Table \ref{tab:NvC6C8CrCr} shows our results for $N_S$, together with the fitted dispersion coefficients, and compares them with theoretical values available in the literature. The availability of just one isotopologue sensitive to odd-$S$ molecular potentials, see Sec.~\ref{Sec:53Cr53Cr}, does not allow for mass-scaling, and thus we cannot determine the number of vibrational levels in $S = 1$, 3, 5. However, by reasonably assuming a monotonic decrease in the number of bound states with increasing $S$, as experimentally shown in the even-$S$ subset and predicted by \emph{ab initio} calculations, we can establish upper and lower bounds for the odd-$S$ channels. As an additional output of our mass-scaled model, relevant for near-threshold atom-atom interactions, we infer the $s$-wave scattering lengths of even-$S$ potentials for all Cr-Cr combinations, which are summarized in Table \ref{tab:scatlengthsCrCr}. Our results, while being consistent with previous ones on $^{52}\mathrm{Cr}$ ($S=6,4$) \cite{PhysRevLett.94.183201}, $^{50}\mathrm{Cr}$ ($S=6$) \cite{PhysRevLett.91.193201} and $^{52}\mathrm{Cr}$-$^{53}\mathrm{Cr}$ ($S=6$)  \cite{PhysRevA.91.011603}, considerably improve their accuracy. We note how an independent measurement of $a_s(S=6)$ from dipolar relaxation of $^{52}$Cr \cite{PhysRevA.81.042716}, if interpreted as the effective low-field scattering length renormalized by the dipole-dipole interaction \cite{PhysRevLett.98.140401}, compares well with our inferred values \cite{ZenodoData} approaching $100.02~a_0$ at zero field. Finally, the value $a_s(S=0)$ of the scattering length associated with the antiferromagnetic ground-state of Cr$_2$, see Table~\ref{tab:scatlengthsCrCr}, is in agreement with the results from spin dynamics measurements on ultracold $^{52}$Cr \cite{PhysRevA.90.043607}.

The only source of theoretical PECs for excited spin states of the chromium dimer is Ref.~\cite{PhysRevA.69.030701} where the complete active space self-consistent field (CASSCF/CASPT2) method is applied, while restricted open-shell Hartree-Fock (ROHF), configuration-interaction second-order perturbation theory (CIPT2) and coupled cluster methods were exclusively employed for the $S=6$ high-spin state. We observe that the CASSCF method systematically underestimates the number of vibrational levels by 10-20\% for the observed even-$S$ states. Using our deduced ranges for odd-$S$ states, we conclude that it also underestimates the number for $S=3$, while it is compatible with our predicted values for $S=5$ and 1, though falling on the lower edge of the ranges. Remarkably, the model of Ref.~\cite{PhysRevA.69.030701} correctly predicts the number of vibrational states of the $S=0$ potential, experimentally determined through laser spectroscopy data in Ref.~\cite{Olivares-Pilon:2024azn}. This may point to an overestimation of the $d$-shell contribution and an underestimation of the $s$-shell one. As far as the $S=6$ state is concerned, we note how CCSDT \cite{NstateS6CrCr} predicts the exact number of states,  RCCSD(T) and CIPT2 predict a number of levels just short of one unit, and ROHF \cite{Cr2Piotr} lies somewhat in between these and the CASSCF prediction. The other states, however, cannot be compared since they are only available for the CASSCF/CASPT2 case. Our fitted $C_6$ coefficient is consistent with the semi-empirical one of Ref.~\cite{PhysRevA.69.030701} and is about 15\% larger than the TDDFT result in Ref.~\cite{Cr2Piotr}. The latter TDDFT calculations also provide a $C_8$ value comparable with the experiment, but similarly underestimated by about 15\%.

\begin{table}[t!]
  \centering
  \caption{The $s$-wave scattering lengths for all the Cr-Cr combinations, in units of Bohr radii. For the $S = 0$ potential, the numbers are calculated from the binding energy of the least bound level estimated in Ref.~\cite{Olivares-Pilon:2024azn}.}
  \label{tab:scatlengthsCrCr}
    \begin{ruledtabular}
      \begin{tabular}{cccccc}
        Isotope  & $^{50}$Cr  & $^{52}$Cr & $^{53}$Cr & $^{54}$Cr & $S$\\ 
        \hline 
        \noalign{\vskip 2mm}
         $^{50}$Cr   & 44.47(3) & -31.49(13) & $-7.6(7)\times 10^3$ & 123.02(14) & \multirow{4}{*}{$6$} \\ 
         $^{52}$Cr   &  &  109.76(11)  & 65.05(4) & 40.91(3) & \\ 
         $^{53}$Cr   &  &  & 39.97(3) & 12.79(5) & \\
         $^{54}$Cr   &  &  &  & -55.5(2) & \\[2mm]
         \hline
        \noalign{\vskip 2mm}
         $^{50}$Cr   & -113.8(4)  & 21.30(4) & $2.22(8)\times 10^3$ & 65.39(4) &  \multirow{4}{*}{$4$} \\ 
         $^{52}$Cr   &  & 56.61(3) & -7.20(8) & 143.4(2) & \\ 
         $^{53}$Cr   &  &  & 132.2(2) & 42.81(3) & \\
         $^{54}$Cr   &  &  &  & -63.2(2) & \\[2mm]
         \hline
        \noalign{\vskip 2mm}
         $^{50}$Cr   & -107.1(13) & -30.8(4) & 80.18(17) & 9.8(2) &  \multirow{4}{*}{$2$} \\ 
         $^{52}$Cr   &  & -9.0(2) & 90.6(2) & 15.0(2) & \\ 
         $^{53}$Cr   &  &  & 11.3(2) & 125.3(5) & \\
         $^{54}$Cr   &  &  &  & 27.13(12) & \\[2mm]
         \hline   
        \noalign{\vskip 2mm}
         $^{50}$Cr   & $41^{+4}_{-4}$ & $31^{+5}_{-5}$ & $168^{+66}_{-32}$ & $28^{+5}_{-6}$ &  \multirow{6}{*}{$0$} \\[2mm]
         $^{52}$Cr   &  & $14^{+6}_{-6}$ & $94_{-9}^{+13}$ & $1^{+9}_{-10}$ & \\[2mm]
         $^{53}$Cr   &  &  & $-4^{+9}_{-13}$& $76_{-7}^{+9}$ & \\[2mm]
         $^{54}$Cr   &  &  &  & $-31^{+16}_{-23}$& \\      
       \end{tabular}
    \end{ruledtabular}
\end{table}

\section{Understanding FR\MakeLowercase{s} and interactions in fermionic $^{53}${C\MakeLowercase{r}}}
\label{Sec:53Cr53Cr}

\begin{figure*}[ht!]
\centering
  \includegraphics[width=\textwidth]
  {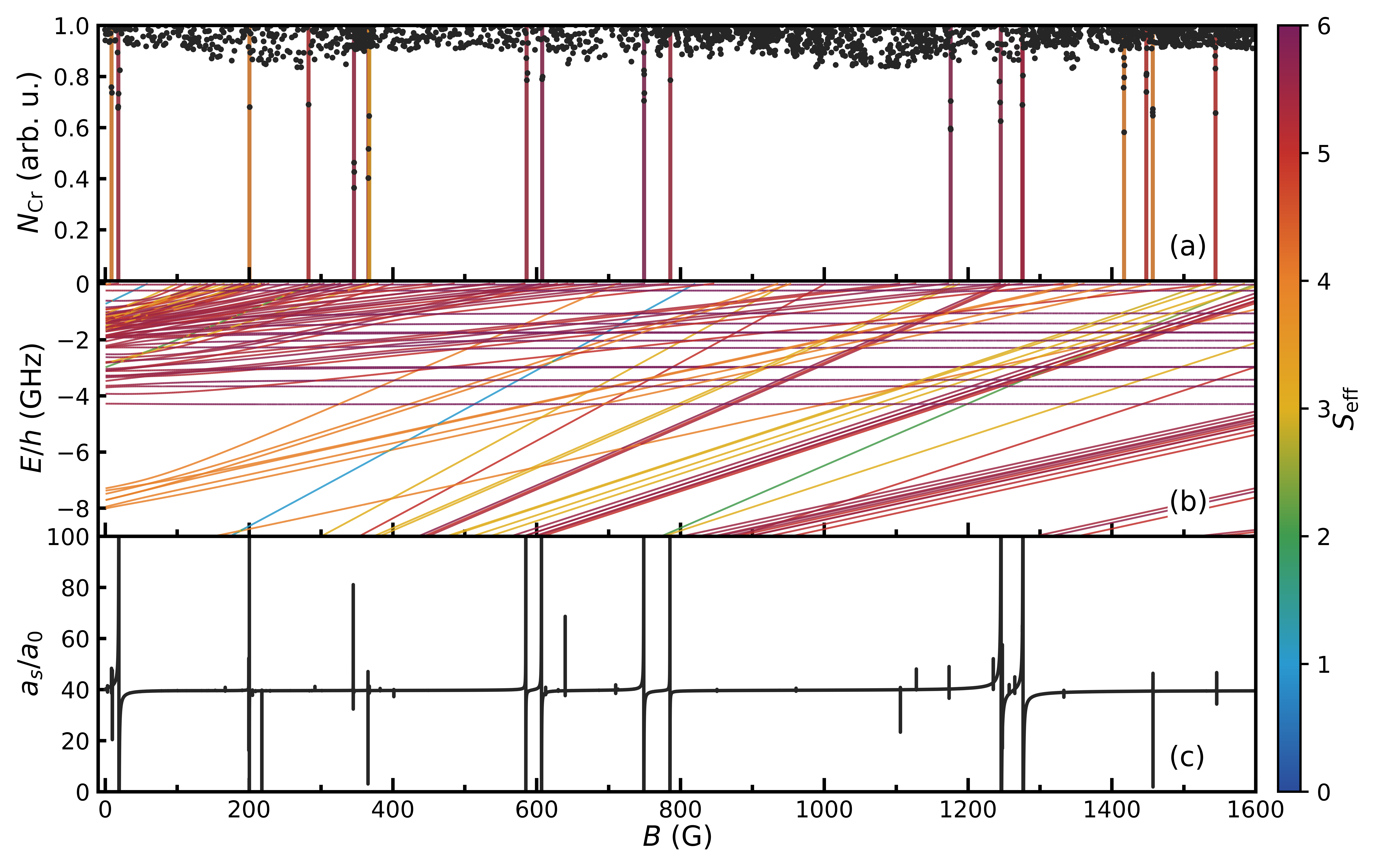}
  \caption{Comparison between the experimental and theoretical Feshbach spectra for the $^{53}\text{Cr}\ket{\text{a}}$-$ ^{53}\text{Cr}\ket{\text{b}}$ system. Panel (a) shows an overview of the low-precision scan, for the $^{53}\text{Cr}\ket{\text{a}}$-$ ^{53}\text{Cr}\ket{\text{b}}$ system, where Feshbach resonances are identified through enhanced atom losses. The vertical lines indicate the resonances positions, and their color indicates the effective spin (see the color bar on the right). Panel (b) shows the energy of the molecular levels (referenced to the scattering channel energy). Bound-state calculations use $M=-8$ and $\ell_{\rm max}=4$. Panel (c) shows the $s$-wave scattering length $a_s/a_0$, where $a_0$ is the Bohr radius, as a function of the magnetic field $B$. Scattering calculations include partial waves up to $\ell_{\rm max}=6$.} 
  \label{fgr:53CrOverview}
\end{figure*}

As a final survey in order of complexity, we investigate the spectrum of $^{53}\text{Cr}$ in the $\ket{\text{a}}$-$\ket{\text{b}}$ channel, in which even under $s$-wave scattering all spin multiplicities are potentially accessible. Moreover, the non-zero nuclear spin ($I=3/2$) opens up spin-exchange coupling, on top of spin-spin, as Feshbach coupling mechanism. These features make the fermionic case significantly more challenging than the bosonic ones, both in terms of assignment and fitting, which motivates our choice to address it only after fixing the even-$S$ mass-scaled potentials on bosonic isotopes.

Our results for the $^{53}\text{Cr}\ket{\text{a}}$-$ ^{53}\text{Cr}\ket{\text{b}}$ from low precision scans are shown in Fig.~\ref{fgr:53CrOverview}(a). We experimentally followed the same procedure used in the previous section and scanned the range from 0 to 1600 G. For each $^{53}$Cr spin state, the atomic densities ranged from $2$ to $3.8\times10^{12}~\mathrm{cm}^{-3}$, with temperatures between $300$ and $400$ nK. Despite the potentially drastic change in the spectrum due to the increased density of molecular levels, our loss spectroscopy data shows a similar density of lines with observed widths comparable to the bosonic case.
On a qualitative level, the most striking difference compared to Fig.~\ref{fgr:50CrOverview} is the appearance of intermediate shades of colors between $S=3$ and $S=6$ used to classify the resonances, reflecting the fact that, as expected, $S$ is not generally a good quantum number in the presence of the hyperfine interaction. The nature of the good quantum numbers depends on the underlying scattering lengths, and, in the case of fermionic Cr considered here, some resonance features are rather associated with atomic-like quantum numbers. Further details on these aspects are provided in Appendix \ref{Appendix:CrCrFRsfermion}.
In analogy with the previous section, Fig.~\ref{fgr:53CrOverview}(b) shows the energy of the molecular levels with the same color code, while Fig.~\ref{fgr:53CrOverview}(c) shows the predicted $B$-field dependence of the $s$-wave scattering length. Since the entrance channel is $s$-wave and interactions only couple it to even partial waves, two-body inelastic losses in the atomic $\ket{\text{a}}$-$\ket{\text{b}}$ spin state, while being energetically allowed, are forbidden by particle exchange symmetry. Hence, the scattering length is real and it diverges in correspondence of the observed resonances.
As evident by comparing Figs.~\ref{fgr:50CrOverview}(b) and \ref{fgr:53CrOverview}(b), the density of states for $^{53}$Cr$_2$ is significantly increased by the presence of the non-zero nuclear spin. All observed features in Fig.~\ref{fgr:53CrOverview}(a) are visible in the theoretically-inferred scattering length in Fig.~\ref{fgr:53CrOverview}(c), with only two exceptions: the FR at 1417\,G, which is allowed but, due to its weakness, appears in Fig.~\ref{fgr:53CrOverview}(b) but is invisible on the scale of Fig.~\ref{fgr:53CrOverview}(c), due to the finite $B$-field step-size, and the FR at 1448 G, which has been assigned via additional close-coupled calculations including incoming $d$-waves for different orbital angular momentum projections $m$. 
In other words, the resonance under consideration belongs to a symmetry block with projection of the total angular momentum $M=-6$, different from that containing the incoming $s$-wave channel ($M=-8$).
\begin{figure*}[t!]
\centering
  \includegraphics[width=\textwidth]
  {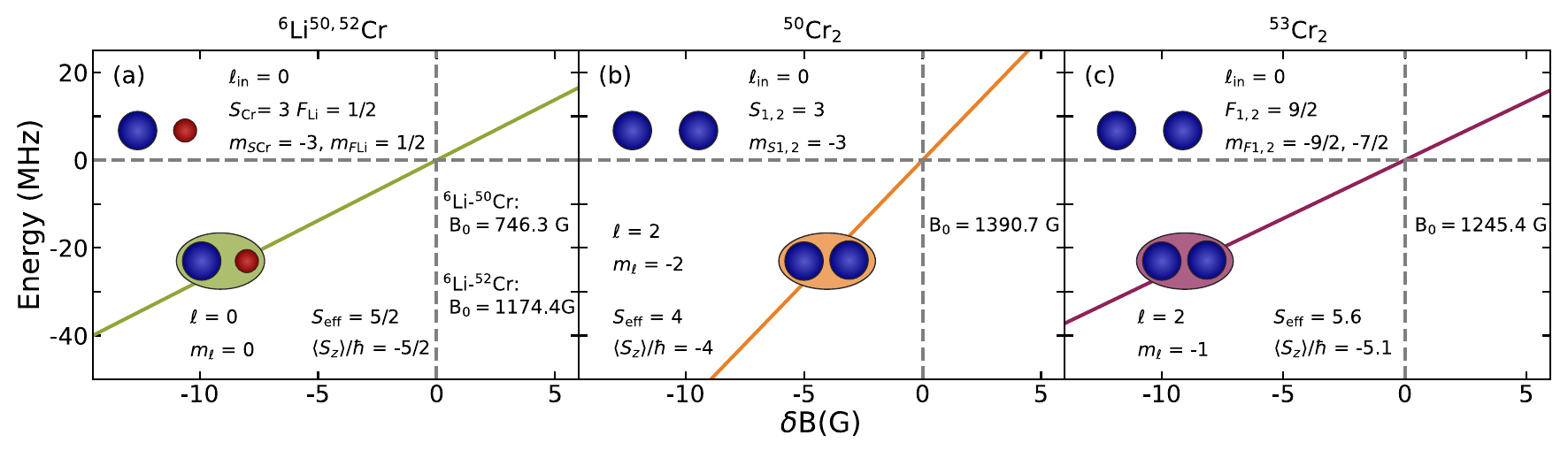}
  \caption{Examples of molecule association pathways for ultracold $^6$Li$^{50}$Cr, $^{50}$Cr$_2$ and $^{53}$Cr$_2$. The energy of the molecular state, referenced to the atomic scattering threshold, is shown as function of the bias magnetic field in the ABSM approximation. 
  The relative slope $\delta \mu$ is the difference between the molecular and atomic moments.
  Weakly-bound $^6$Li$^{50}$Cr and $^{50}$Cr$_2$ follow Hund's coupling case (b), while $^{53}$Cr$_2$ shows atomic-type spin correlation due to the coincidental quasi-degeneracy of $a_s(6)$ and $a_s(5)$. The pathway shown in panel (b) is not the best possible but is the one used as a proof for molecule association, see Fig.~\ref{fgr:lineshape}.} 
  \label{fgr:molFormation}
\end{figure*}

We can explain the observed features in terms of resonant fields by assigning all resonances and fitting our quantum collisional model. After fixing the FRs by mass-scaling the even-$S$ interactions, we are left only with the odd-$S$ scattering lengths as free fitting parameters.
\begin{table}[b!]
  \centering
  \caption{The $s$-wave scattering lengths, in units of the Bohr radius, for $^{53}$Cr-$^{53}$Cr across electron spin multiplicities. The values are obtained from our model, optimized against the resonance spectra of \emph{both} bosonic and fermionic chromium isotopes. The best-fit dispersion coefficients are $C_6=732.61~ E_h a_0^6$ and $C_8=3.972\times10^4 ~ E_h a_0^8$, consistent with those obtained earlier in this work from a fit of the bosonic isotopes only.}
  \label{tab:scatlengthsCr53}
    \begin{ruledtabular}
      \begin{tabular}{cccccccc}
        $S$  & 6  & 5 & 4 & 3 & 2 & 1 & 0\\ [1mm]
        \hline 
         \noalign{\vskip 1mm}
         $a_s$   & 39.92(8) & 40.78(7) & 132.0(4) & 46.9(1.8) & 11.3(3) & & $-4^{+9}_{-13}$ \\
          \noalign{\vskip 1mm}
       \end{tabular}
    \end{ruledtabular}
\end{table}

We include bound levels with up to $\ell=6$ rotational quantum number and obtain an unambiguous assignment, with $S=5$ explaining most of the observed features, and $S=3$ responsible for the remaining subset. The resulting values of the scattering lengths are summarized in Table~\ref{tab:scatlengthsCr53}. Interestingly, the similarity of the scattering lengths $a_s(5)$ and $a_s(6)$ explains the narrow character of the $s$-wave spin-exchange resonance at about 1544\,G, and the emergence of atomic-like hyperfine coupling ($f_a$,$f_b$) as a result of hyperfine interactions. The $S = 4$ states indeed tend to remain nearly pure, while the $S=2$ and $S=3$ states are rarely pure, but they do influence the position of some resonances. As an indirect confirmation of this, Table \ref{tab:scatlengthsCr53} shows that $a_s(3)$ is rather well constrained by the measurements, with an uncertainty on the order of $a_0$. 
Note that the even-$S$ scattering lengths values in Table \ref{tab:scatlengthsCr53} are not identical to those listed in Table \ref{tab:scatlengthsCrCr}, as they are now obtained from a simultaneous fit to the bosonic and fermionic data.
Finally, as for the case of $^{50}$Cr, we test the magnetic-field width of a selected FR at 1245~G, for which we measure $\Delta_{\text{exp}}=660(60)$~mG, see Appendix \ref{app:UhlOrn}, in very good agreement with the theory value $\Delta_{\text{th}}=705$~mG.

We highlight that our final level of accuracy for the Cr$_2$ system has been reached neglecting second-order spin-orbit interactions, which are present in $S>0$ but unknown. Their strength generally depends on the $S$ quantum number and, for the comparably heavy Cr$_2$, they are expected to compete at short range with direct electron spin-spin coupling, from which they are experimentally indistinguishable. Hence, it is hard to assess the accuracy of theoretical resonant magnetic fields and widths, and the feasibility of a complete assignment \emph{a priori} for such a system. The successful case of Cr$_2$ is thus encouraging for the investigation of other complex diatomics through experimental strategies similar to those employed herein. 

Given the complex structure of the bound-state energy spectrum, we have analysed the nearest-neighbour level-spacing distribution to investigate possible signatures of quantum chaotic behaviour. Following Ref. \cite{PhysRevA.93.022703}, we have fitted the distribution using the Brody distribution, which interpolates between the Poisson and Wigner-Dyson limits. The resulting distribution is close to the Poisson limit, indicating a limited degree of level repulsion. This behaviour can be expected given the approximate conservation of various quantum numbers, particularly $\ell$ and $m_f$, preventing clear signatures of quantum chaos from emerging in the overall spectrum.

\section{Molecule assembly and implications for ultracold L\MakeLowercase{i}-C\MakeLowercase{r} and C\MakeLowercase{r} gases}
\label{Sec:Prospects}

\begin{figure}[t!]
\centering
  \includegraphics[width=\columnwidth]
  {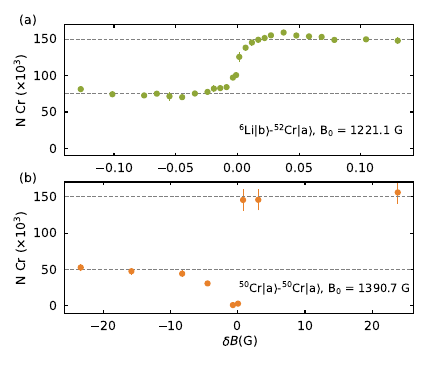}
  \caption{Atomic losses signals through two Feshbach resonances of $^6$Li$^{52}$Cr (a) and $^{50}$Cr$_2$ (b). The measurement on panel (a) is taken scanning the final value of a magnetoassociation ramp of 20 ms starting from 200 mG on the negative scattering length side of the Feshbach resonance. The imaging is performed after 5 ms of wait time at the target field. The densities of the parent atomic sample are $0.9 \times 10^{12} \text{cm}^{-3}$ for Li and $0.2 \times 10^{12} \text{cm}^{-3}$ for Cr, while the sample for the measurement of panel (b) was a Bose-Einstein condensate (see Ref.~\cite{LiCrMixtures}). For panel (b) we performed a 50 ms ramp from around 1414 G to the target field and after waiting 300 ms we performed the imaging.} 
  \label{fgr:lineshape}
\end{figure}

Both heteronuclear Li-Cr and homonuclear Cr systems are favorable for molecule assembly. LiCr dimers of both fermionic and bosonic statistics are produced via magnetoassociation, allowing to investigate the impact of quantum statistics, under identical resonant scattering properties \cite{CrFeshbachMols}. The employed association pathways for fermionic $^6$Li$^{50,52}$Cr are shown in Fig.~\ref{fgr:molFormation}(a), to be compared with the case of bosonic $^6$Li$^{53}$Cr in Ref. \cite{PRXQuantum.5.020358}. We magnetoassociate molecules with a conversion efficiency of about 50\% in the single beam ODT in the low end of density range explored in this paper, where the details on the association mechanism will be the subject of Ref. \cite{CrFeshbachMols}. The corresponding loss signal is shown in Fig.~\ref{fgr:lineshape}(a) in analogy with Ref. \cite{PRXQuantum.5.020358}. 
Furthermore, thanks to the isolated Gauss-wide resonances both in bosonic and fermionic isotopes, we produce transition-metal dimers at ultralow temperatures for the first time. In Figs.~\ref{fgr:molFormation}(b) and (c) we show a plot of the molecule association pathway for the new homonuclear molecules $^{50,53}$Cr$_2$. In particular, we show losses induced by magnetoassociated $^{50}$Cr$_2$ from a Bose-Einstein condensate~\cite{LiCrMixtures}, see Fig.~\ref{fgr:lineshape}(b). The association mechanisms, especially efficient for the fermionic counterpart, will be reported with further details in Ref.~\cite{CrFeshbachMols}. Ground-state Cr$_2$ dimers would be interesting for ultracold chemistry and precision measurements, in particular variation of the proton-to-electron mass ratio, thanks to the large sensitivity in the $S=0$ state (up to about 4000 cm$^{-1}$ for relative change of the mass ratio of one unity \cite{PhysRevLett.100.043202}) and the potential for quasi-degenerate levels of different $S$. Remarkably, one-photon spectroscopy has been already performed on this system, with quantum number assignment even for excited states~\cite{ruhrig_photoassociation_2016}.

The results obtained in the present work allow us to draw some conclusions about future applications on Li-Cr mixtures and Cr gases.
$^6$Li-$^X$Cr spin-exchange sextet resonances, featuring similar parameters (width, background scattering length) across the different isotopic combinations, see Table \ref{tab:FRsLiCr}, appear as an almost unique playground where to investigate the effect of distinct quantum statistics under identical resonant scattering conditions, immune to two-body losses \cite{Baroni2024}.
Moreover, interesting scenarios are expected for the $^6$Li-$^{50}$Cr mixture at the three-body level and beyond, given the presence of Li-Cr interspecies resonances falling within the BCS-BEC crossover region of $^6$Li \cite{PhysRevLett.92.040403,Bartenstein2004,Zwierlein2004,Bourdel2004} in correspondence of BEC, unitary and BCS regimes of either  $^{6}\text{Li}\ket{\text{a}} - \ket{\text{b}}$ or $\ket{\text{a}}-\ket{\text{c}}$ combinations. As far as Cr homonuclear gases are concerned, we note how $^{50}$Cr, albeit similar to $^{52}$Cr
in terms of Feshbach coupling mechanisms, offers a different playground, with a smaller background scattering length and three main FRs. Among those, one is centered at about 5\,G, a sufficiently low field to allow to tune the magnetic field (and thus the dipoles) orientation compared to a given optical trap geometry. Another note on bosonic chromium concerns the $^{54}$Cr isotope, unresolved at the laser cooling stage, for which our mass-scaled model predicts a negative $s$-wave scattering length in its lowest Zeeman level, see Table \ref{tab:scatlengthsCrCr}. While this does not prevent the production of ultracold samples, it would preclude stable Bose-Einstein condensates without a resonant control of the interactions.

\section{Conclusions and outlook}
\label{Sec:Conclusions}

In conclusion, we have investigated three chromium-bearing diatomics with increasing level of complexity: $^{6}$Li$^{X}$Cr, $^{50,52}$Cr$_2$, and $^{53}$Cr$_2$. For all of them we have performed high-precision Feshbach spectroscopy on the parent atomic mixtures, from which we successfully built complete quantum collisional models for their interactions. Our models for the first two systems can be mass-scaled, allowing us to determine the number of vibrational levels across different spin states. On the one hand, LiCr only possesses two spin-multiplicities and the most recent \emph{ab initio} coupled-cluster results \cite{PRXQuantum.5.020358} match our experimental findings. On the other, Cr$_2$ features 7 spin-multiplicities, out of which, combining our results with the already known $S=0$ ground state, we provide the exact number of vibrational levels for all even-$S$ states.
The predictions from multiconfiguration calculations (CASSCF/CASPT2) appear to underestimate the number of vibrational levels by 10\% to 20\% , suggesting a systematic underestimation of the associated potential well depths. Nonetheless, the accuracy of the initial theoretical PECs was sufficient to achieve a complete assignment for the observed resonances.
This should not be taken for granted, as spin-spin couplings, unlike the spin-exchange interaction, depend on the shape of the PEC over a broad range of internuclear distances and cannot be simply described by purely short-range parameters. For the $S=6$ potential, we compared predictions from several \emph{ab initio} methods and found the best agreement with the coupled-cluster results \cite{Cr2Piotr,NstateS6CrCr}.
In addition, we compared the fitted dispersion coefficients of LiCr and Cr$_2$ systems, finding best agreement with coupled-cluster and semi-empirical results, respectively. Furthermore, our work allows to fix the $s$-wave scattering lengths of both spin multiplicities of all Li-Cr isotopic combinations, of even-$S$ states for all Cr-Cr combinations, plus of all but $S=1$ in fermionic $^{53}$Cr. We used the acquired knowledge to assemble Cr-bearing diatomics with quantum state and isotopologue selectivity, as well as to propose potential applications of LiCr and Cr gases for many-body physics studies.

Our work shows how state-of-the-art \emph{ab initio} calculations and fitting of high-precision FR spectra can guide magneto-association and optical ground-state transfer of ultracold molecular species of considerably higher complexity than bi-alkalis. Our findings suggest the experimental feasibility of ultracold molecule production of targeted chemical species for quantum simulation, ultracold chemistry and precision measurements. As a specific example, it is worth mentioning YbCr which has been proposed for next-generation charge-parity violation searches via the electron electric dipole moment (eEDM) and magnetic quadrupole moment (MQM) with selected isotopologues \cite{ciamei_ultracold_2026}. These considerations could be extended for instance to polar paramagnetic molecules containing europium \cite{PhysRevA.90.022514,PhysRevA.98.032704}, which has been recently Bose-condensed and shown to feature suitable FRs \cite{PhysRevLett.129.223401}: while Eu is expected to behave similarly to Cr at the molecule assembly level, it could allow for high sensitivity to the nuclear Schiff moment (NSM) on its stable octupole-deformed $^{153}$Eu isotope \cite{PhysRevC.111.055501, nima_limit_2026,chen_relativistic_2024}.

\section{Acknowledgments}
\label{Sec:Acknowledgements}

We thank Prof.~P.~\.{Z}uchowski and Prof.~M.~Tomza for useful discussions. This work was supported by the European Research Council under Grant No. 101220504(COMPASS), by the Italian Ministry of University and Research under the PRIN2022 project No. 20227F5W4N and, co-funded by the European Union — NextGenerationEU, under the ‘Integrated infrastructure initiative in Photonic and Quantum Sciences’ I-PHOQS (CUP B53C22001750006), and the Young Researcher Grants MSCA 0000042 (PoPaMol, fellowship to A.C.). We also acknowledge support from the project 23FUN04 COMOMET, which has received funding from the European Partnership on Metrology, co-financed by the European Union’s Horizon Europe Research and Innovation Programme and by the Participating States (Funder ID: 10.13039/100019599). Numerical calculations were performed on the SCaPhyR (Serveur de Calcul pour la Physique Rennaise) cluster.

\section{Data availability}
\label{Sec:Repository}
The data that support the findings of this article are
openly available \cite{ZenodoData}.

\newpage
\appendix

\section{Details on Li-Cr FRs}
\label{Appendix:LiCrFRs}

This appendix summarizes the main experimental and theoretical properties of the Li-Cr Feshbach resonances considered in Sec.~\ref{Sec:LiCr}. The relevant parameters are collected in Table \ref{tab:FRsLiCr}, where we report the resonance positions determined experimentally $B_{0,\text{exp}}$, together with the corresponding theoretical predictions.

For a generic angular momentum $\vec J$ with mean-square value $\langle \vec{J}^{\,2} \rangle$, we find it convenient to define an effective quantum number $J_{\rm eff}$ through the relation $J_{\rm eff}(J_{\rm eff}+1)\hbar^2 =\langle \vec{J}^{\, 2} \rangle $. We report in particular the effective total spin angular momentum $f_{\rm eff}$, the effective electron spin $S_{\rm eff}$ and the effective nuclear spin $I_{\rm eff}$ along with their average projections $\langle f_z\rangle $, $\langle S_z\rangle$ and $\langle I_z \rangle$.
Conservation of $f_z$ is broken only by the anisotropic spin-spin interaction, which is relatively weak in this system, such that the corresponding quantum number $m_f$ remains essentially well defined. 

Moreover, the difference between the sextet and octet scattering lengths leads to energy separations between corresponding sextet and octet ro-vibrational levels that exceed the hyperfine splitting. Consequently, despite the hyperfine interaction coupling molecular states of the sextet and octet potentials, the total electronic spin $S$ remains a good quantum number in most cases.
Furthermore, at the experimental fields, the Zeeman interaction uncouples $\vec S$ and $\vec I$, such that the spin projections $M_S$ and $M_I$ are also good quantum numbers for most features. To a very good approximation, all molecular states have $\ell \approx 0$ and are coupled to the incoming $s$-wave channel by the spin-exchange interaction.  
 
Note that some of the features listed in Table~\ref{tab:FRsLiCr} are observed in collisions starting from excited levels of the hyperfine manifold thus not immune to 2-body inelastic decay onto lower-lying states. In these cases, we parametrize $s$-wave collisions {\it exactly} introducing the complex scattering length and its pole decomposition near an isolated resonance as \cite{Hutson_2007,Simoni_2021}: 
\begin{equation}\label{eq:complex_asc}%--------------------------------
a = a_{\rm bg} \left[ 1-  \frac{\Delta}{(B-B_0)+ i \, \Gamma_{\rm in} / 2} \right] ,
\end{equation}%---------------------------------------------------------
where $\Gamma_{\rm in}$ is the magnetic decay width, $\Delta$ the generally complex resonance magnetic width, and the non-resonant contribution $a_{\rm bg}$ is generally a slowly varying function of $B$. In cases where two or more resonances significantly overlap, the spectrum is modeled using the generalization of equation \eqref{eq:complex_asc} to overlapping features \cite{Simoni_2021}: 
\begin{equation}\label{eq:complex_asc_overlap}%--------------------------------
a = a_{\rm bg} \left[ 1-  \sum_k \frac{\Delta_{k}}{(B-B_{0,k})+ i \, \Gamma_{{\rm in},k} / 2} \right] .
\end{equation}%

In addition to the resonances discussed in Sec.~\ref{Sec:LiCr}, we also include the experimentally observed features in the combination Li$\ket{\text{b}}$-Cr$\ket{\text{a}}$. Although these resonances are not required to constrain the mass-scaled model, measurement of their isotopic shifts provides an additional experimental benchmark of the number of bound states supported by the sextet potential.

\section{Details on $^{50}$Cr and $^{52}$Cr FRs}
\label{Appendix:CrCrFRs}

This appendix provides an overview of the main experimental and theoretical properties of the $^{50}$Cr-$^{50}$Cr Feshbach resonances investigated in Sec.~\ref{Sec:50Cr50Cr}.

A summary of the corresponding quantities is given in Table \ref{tab:FRsCr50}. For each resonance, we list the measured $B_{0,\text{exp}}$ and theoretically predicted $B_{0,\text{th}}$ magnetic field positions, the magnetic field widths $\Delta_{\text{th}}$ obtained from the theoretical model and the average quantum numbers associated with the corresponding molecular states. Specifically, we report the effective angular momentum $\ell_{\text{eff}}$, the effective spin $S_{\text{eff}}$ and its projection $\langle S_z \rangle$.

\begin{figure*}[ht!]
\centering
  \includegraphics[width=\textwidth]
  {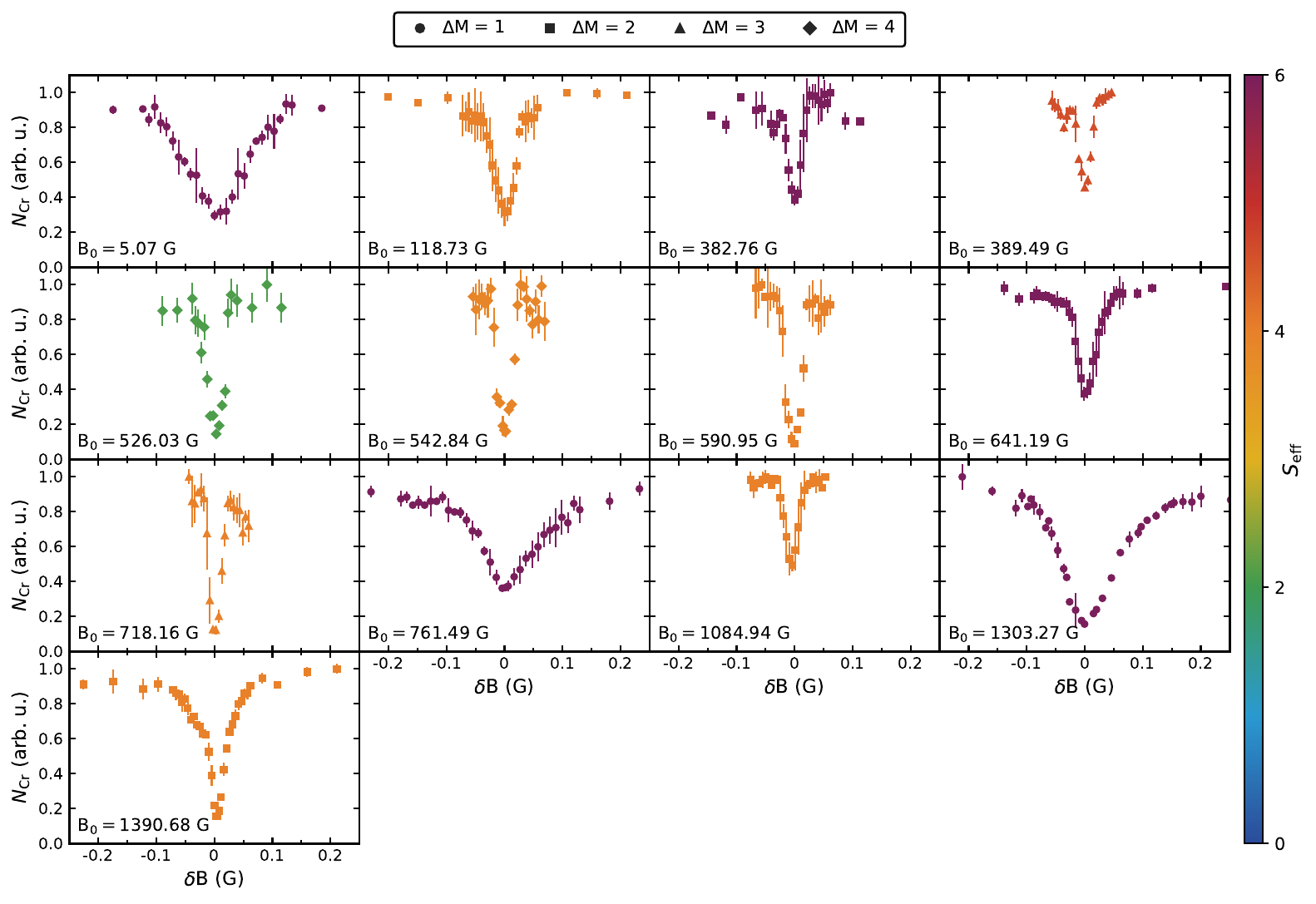}
  \caption{Fine scans of the $^{50}$Cr-$^{50}$Cr Feshbach resonances. The color scale reflects the effective spin, see color bar on the right. The markers identify the different $\Delta M$, see legend above.} 
  \label{fgr:50CrZooms}
\end{figure*}

The effective spin is generally a well-defined quantum number, as shown in Table~\ref{tab:FRsCr50} and Fig.~\ref{fgr:50CrZooms}, which presents fine scans of the $^{50}$Cr$\ket{\text{a}}$-$^{50}$Cr$\ket{\text{a}}$ channel. Exceptions occur at specific magnetic fields where states with different values of $S$ and $\ell$ become accidentally nearly degenerate. In such cases, the spin-spin interaction can mix these states, causing the corresponding expectation values to deviate from integer numbers. This effect is particularly evident for the resonance at 389 G, where the reddish color indicates a superposition of states with different spin quantum numbers. Different marker shapes are used in Fig.~\ref{fgr:50CrZooms} to distinguish resonances according to their $\Delta M$ values. Notably, the resonances that appear widest have $\Delta M = 1$.

\begin{figure*}[h!]
\centering
  \includegraphics[width=\textwidth]
  {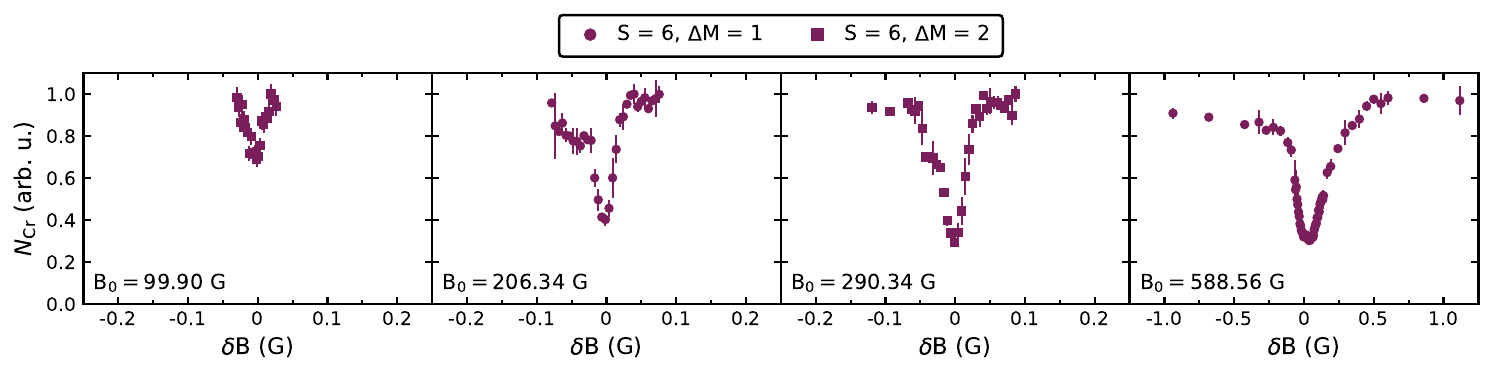}
  \caption{Fine scans for the $^{52}$Cr-$^{52}$Cr Feshbach resonances first reported in Ref. \cite{PhysRevLett.94.183201} retaken with our ultracold sample. All the resonances are induced from the $S=6$ potential, with different $\Delta M$.} 
  \label{fgr:52CrZooms}
\end{figure*}

We also report the $^{52}$Cr-$^{52}$Cr Feshbach resonances measurements obtained with colder samples than in Ref.~\cite{PhysRevLett.94.183201}, reporting the adjusted resonance positions in Table~\ref{tab:52CrComparison}. The corresponding loss spectra are shown in Fig.~\ref{fgr:52CrZooms}. In this case all the reported resonances are induced by the $S=6$ state while having different $\Delta M$. 

\begin{table}[t!]
  \centering
  \caption{Comparison of our experimental and theoretical $^{52}$Cr resonance positions with the experimental values reported by the Stuttgart group \cite{PhysRevLett.94.183201}.}
  \label{tab:52CrComparison}
    \begin{ruledtabular}
      \begin{tabular}{ccc}
        $B_{0, {\rm exp}}$ (G) & $B_{0, {\rm exp}}$ (G) & $B_{0,{\rm th}}$ (G) \\[1mm]
        This work & Ref. \cite{PhysRevLett.94.183201} & This work\\
        \hline
      \noalign{\vskip 2pt}
         98.898(20)  & 98.90(14) & 99.2(3) \\ 
         206.338(20) & 205.80(14) & 207.2(6) \\ 
         290.339(20) & 290.30(14) & 290.9(6) \\
         588.560(40) & 589.10(14) & 589.7(7) \\
       \end{tabular}
    \end{ruledtabular}
\end{table}

\section{Details on $^{53}$Cr FRs}
\label{Appendix:CrCrFRsfermion}

This appendix provides an overview of the main experimental and theoretical properties of the $^{53}$Cr$\ket{\text{a}}$-$^{53}$Cr$\ket{\text{b}}$ Feshbach resonances investigated in Sec.~\ref{Sec:53Cr53Cr}, summarized in Table \ref{tab:FRsCr53}. For each resonance, we list the experimentally determined $B_{0,\text{exp}}$ and theoretical $B_{0,\text{th}}$ magnetic field positions, the magnetic field widths $\Delta$ obtained from the theoretical model and the quantum numbers characterizing the molecular states involved.

Compared with alkali atoms, Li-Cr, or bosonic Cr–Cr systems, the present case is more complex to analyze due to the interplay of high spin multiplicity, hyperfine interactions, and stronger anisotropic dipole–dipole interactions.
We consider two versions of a molecular-like representation based on the $ \{ \ell S I  \}$ quantum numbers. In the first, the electron and nuclear spins are coupled to
form the hyperfine quantum number, yielding the basis states $| \ell m S I  f m_f \rangle$.
In the second, each angular momentum is independently quantized along the direction of the external magnetic field, yielding $| \ell m S M_S I  M_I \rangle$.

We also consider two atomic-like representations based on the $\{\ell,f_a,f_b\}$ quantum numbers, corresponding to the coupled and uncoupled basis states $|\ell m f_a f_b f m_f\rangle$ and $|\ell m f_a m_{f_a} f_b m_{f_b}\rangle$, respectively.
These states are implicitly symmetrized with respect to the exchange of identical atoms.
In order to identify the best assignment, we analyze the projections of each numerically calculated molecular state onto our four alternative bases. We then report in the table only those quantum numbers in the selected representation whose purity exceeds a threshold of 0.9, arbitrarily fixed.

\begin{figure*}[ht!]
\centering
  \includegraphics[width=\textwidth]
  {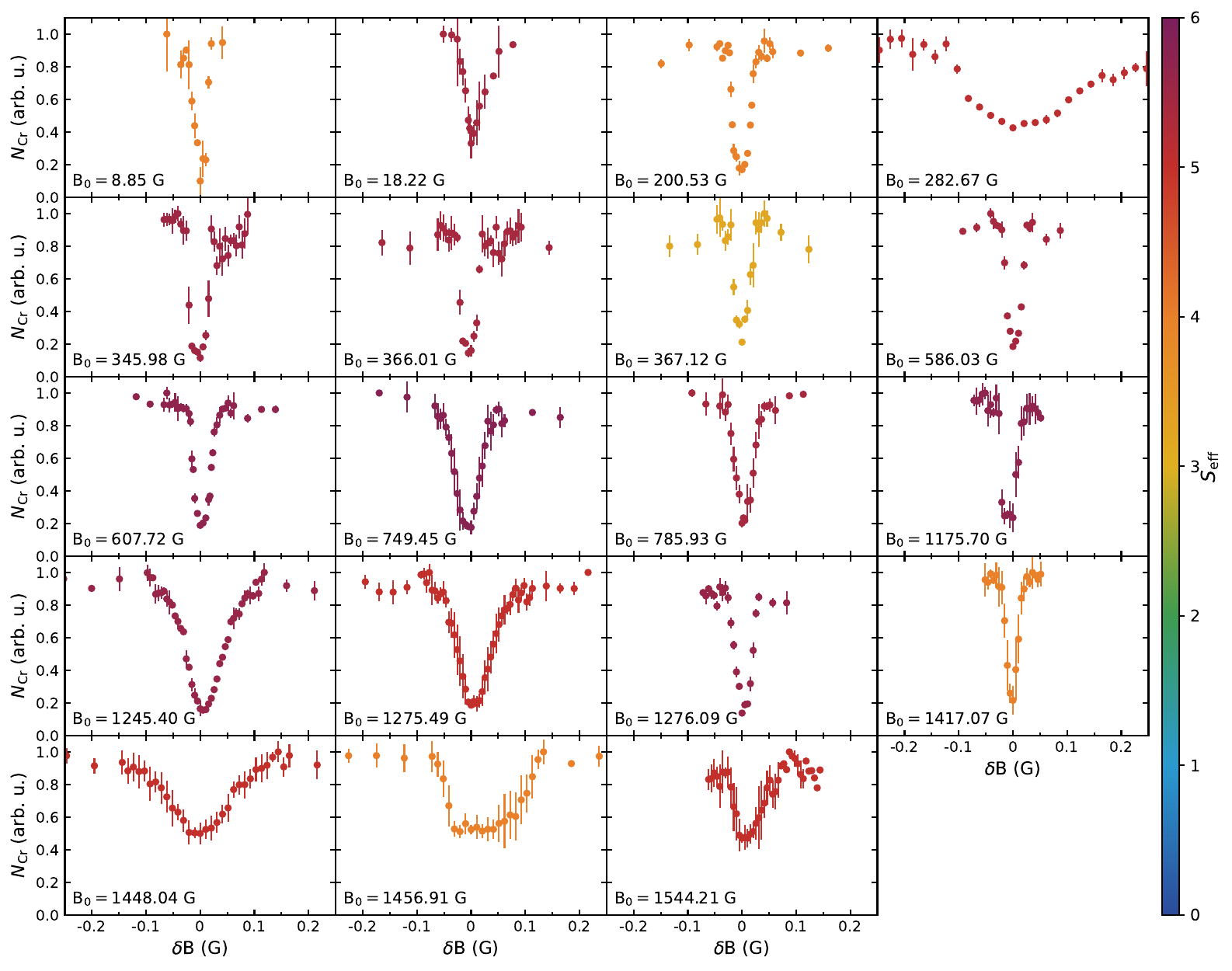}
  \caption{Fine scans of the $^{53}$Cr-$^{53}$Cr Feshbach resonances. The color scale reflects the effective spin, see color bar on the right.} 
  \label{fgr:53CrZooms}
\end{figure*}

Fig.~\ref{fgr:53CrZooms} shows fine scans of the $^{53}$Cr$\ket{\text{a}}$-$^{53}$Cr$\ket{\text{b}}$ FRs, with the color scale representing the effective spin quantum number.

Some remarks are in order.
The rotational angular momentum $\ell$ is conserved in the absence of anisotropic interactions and remains essentially a good quantum number for nearly all resonances, even in the presence of the dipole–dipole interaction. The total hyperfine angular momentum $f$, by contrast, is exactly conserved only in the absence of both a magnetic field and anisotropic interactions, and remains an approximately good quantum number for low-field features, as shown in the table.

The emergence of $f_a$ and $f_b$ as good quantum numbers stems from a remarkable similarity between the numerical values of the scattering lengths associated with the adiabatic potentials $S=5$ and $6$. As a result, the corresponding ro-vibrational levels, counted down from dissociation, are nearly degenerate. This near degeneracy allows the hyperfine interaction to efficiently mix these levels, favoring a representation based on the atomic quantum numbers $f_a$ and $f_b$. At large magnetic fields, the hyperfine atomic angular momenta $\vec{f}_{a}$ and $\vec{f}_{b}$ precess independently about the field direction, making $m_{fa}$ and $m_{fb}$ good quantum numbers.

Note that hyperfine mixing requires that, for a given value of $m_f$, states with $S=5$ and $S=6$ exist with the {\it same} value of $M_S$. In this case, the corresponding levels remain approximately degenerate as a function of $B$ and can therefore mix.
A notable example in which mixing {\it does not} occur is the only resonance observed with $\ell = 0$, located near $1544$~G . According to the model, this resonance is purely $S=5$. This purity arises from the fact that only two molecular states with $m_f=-8$ are compatible with the nuclear-exchange symmetry, which requires $S+I+\ell$ to be even. These are the states $| \ell=0 ~ m=0 ~ S=6 ~ M_S=-6 ~ I=2  ~ M_I=-2 \rangle$ and $| \ell=0 ~ m=0 ~ S=5 ~ M_S=-5 ~ I=3 ~ M_I=-3 \rangle$. At the experimental magnetic field, the Zeeman interaction clearly lifts the ro-vibrational degeneracy, preventing the two states from mixing.

Finally, the ro-vibrational levels supported by the $S~<~5$ potentials are in general well separated on the scale of the hyperfine coupling and therefore tend to retain a good $S$ character. Of course, given the high density of interacting levels, there are also cases in which the aforementioned quantum numbers are mixed and none of them emerges as a sufficiently pure quantum number.

Some resonances have also been identified in the $^{53}$Cr$\ket{\text{b}}$-$^{53}$Cr$\ket{\text{c}}$ collision channel, at 1273.8(4), 1274.6(3) and 1275.1(3) G, with their positions well reproduced by our model.

\begin{table*}[t!]
  \centering
  \caption{Experimental magnetic-field positions of the employed Feshbach
  resonances for $^6$Li-$^X$Cr, 
  together with the corresponding theoretical positions, the real and imaginary parts of the background scattering length $a_{\rm bg}$ and of the resonance width $\Delta_{\rm th}$, the decay width $\Gamma_{\rm in}$, and the average quantum numbers (see text).}
  \label{tab:FRsLiCr}
  \begin{ruledtabular}
    \begin{tabular}{cccccccccccccc}
      Channel
      & $B_{0, \rm exp}$(G)
      & $B_{0, \rm th}$(G)
      & $a_{\rm bg,r}$($a_0$)
      & $a_{\rm bg,i}$($a_0$)
      & $\Delta_{{\rm th}, r}$(mG)
      & $\Delta_{{\rm th}, i}$ ($\mu$G)
      & $\Gamma_{\rm in}$(mG)
      & $f_{\rm eff}$
      & $\langle f_z\rangle/\hbar$
      & $S_{\rm eff}$
      & $\langle S_z\rangle/\hbar$
      & $I_{\rm eff}$ 
      & $ \langle I_z\rangle/\hbar$
      \\[1mm]
      \hline
      \noalign{\vskip 2pt}
      \multicolumn{14}{l}{$^6$Li-$^{50}$Cr} \\
      \hline
      \noalign{\vskip 2pt}
      $|\rm{a a} \rangle$
      & 124.089(20) & 124.128 & 46.6 &  0 & 1.1 & 0 & 0
      & 4.2 & -2.5 & 3.5 & -1.5 & 1 & -0.98 \\

      $|\rm{a a}\rangle$
      & 746.318(20) & 746.383 & 46.9 &  0 & 504 & 0 & 0
      & 2.8 & -2.5 & 2.5 & -2.5 & 1 & 0.012 \\

      $|\rm{b a}\rangle$
      & 793.065(20) & 793.114 & 46.9 & $3.1\times10^{-6}$ & 514 & -0.13 & $1.33\times10^{-4}$
      & 3.5 & -3.5 & 2.5 & -2.5 & 1 & -0.99 \\[1mm]

      \hline
       \noalign{\vskip 2pt}
      \multicolumn{13}{l}{$^6$Li-$^{52}$Cr} \\
      \hline
      \noalign{\vskip 2pt}
      $|\rm{a a}\rangle$
      & 160.454(20) & 160.526 &42.8 & 0 & 0.58 & 0 & 0
      & 3.7 & -2.5 & 3.0 & -1.5 & 1 & -0.99 \\

      $|\rm{aa}\rangle$
      & 1174.425(20) & 1174.205 & 43.1 &0 & 491 &0 & 0
      & 2.8 & -2.5 & 2.5 & -2.5 & 1 & 0.01 \\

      $|\rm{b a}\rangle$
      & 1221.057(20) & 1220.826 &43.1 & $1.4\times10^{-6}$ & 498 & -0.058 & $5.77\times10^{-5}$
      & 3.5 & -3.5 & 2.5 & -2.5 & 1 & -0.99 \\[1mm]

      \hline
      \noalign{\vskip 2pt}
      \multicolumn{13}{l}{$^6$Li-$^{53}$Cr} \\
      \hline
    \noalign{\vskip 2pt}
      $|\rm{b b}\rangle$
      & 167.737(20) & 168.145 & 40.6 & $1.9\times10^{-3}$ & 6.7 & -39 & $2.8$
      & 4.7 & -4.0 & 3.5 & -1.7 & 2.4 & -2.3 \\

      $|\rm{a a}\rangle$
      & 1413.886(20) & 1414.140 & 41.4 & 0 & 475 & 0 & 0
      & 4.2 & -4.0 & 2.5 & -2.5 & 1.9 & -1.5 \\

      $|\rm{ba} \rangle$
      & 1460.933(20) & 1461.212 & 41.4 &  $9.7\times10^{-7}$& 480 & -0.041 & $4.31\times10^{-5}$
      & 5.0 & -5.0 & 2.5 & -2.5 & 2.5 & -2.5 \\

    \end{tabular}
  \end{ruledtabular}
  \vskip -13pt
\end{table*}

\begin{table*}[h!]
  \centering
  \caption{Experimental and theoretical magnetic-field positions, theoretical background scattering length $a_{\rm bg}$ and width $\Delta$, and average quantum numbers of the observed $^{50}$Cr-$^{50}$Cr Feshbach resonances.}
  \label{tab:FRsCr50}
    \begin{ruledtabular}
        \begin{tabular}{ccccccc}
            $B_{0, \text{exp}}$(G) & $B_{0, \text{th}}$(G) & $a_{\rm bg}(a_0)$ & $\Delta_{\rm th}$(mG) & $ \ell_{\rm eff} $ &  $S_{\rm eff} $  & $\langle S_z \rangle /\hbar$ \\ 
            \hline  
            \noalign{\vskip 2mm}
            5.066(20) & 5.505 & 46.6 & 442 & 1.9  & 6.0 & -4.8   \\ 
            118.731(20)& 118.969 & 47.4 & 130&  2.0 & 4.0 & -4.0 \\
            382.762(20)  & 382.852 &47.7  & 22.4  & 4.0 & 6.0 & -4.0 \\
            389.489(20) & 389.468 &47.7 & 6.60 & 5.4 & 4.6 & -3.3 \\
            526.03(10) & 525.946 &47.7 & 0.12 & 4.0 & 2.1 & -2.0\\
            542.84(10) & 543.231 &47.7 & 0.31 & 4.0 & 3.9 & -2.0 \\
            590.95(10) & 591.516 &47.7 & 0.78 &6.0 & 4.0 & -4.0 \\
            641.189(20) & 641.087 &47.8 & 290 & 2.0 & 6.0 & -4.0\\
            718.16(10) & 718.349 &47.8 & 6.5  &4.0& 4.0& -3.0\\
            761.486(20) & 759.945 &47.9 & 1100 & 4.0  & 6.0 & -5.0 \\
            1084.936(20) & 1086.192 &47.8 & 17.8 & 4.0 & 4.0 & -4.0\\
            1303.271(20) & 1303.935 &47.8 & 570 & 2.0  & 6.0 & -5.0\\
            1390.683(20) & 1388.635 &47.8 & 127  & 2.0 & 4.0& -4.0\\
       \end{tabular}
    \end{ruledtabular}
 \vskip -12pt
\end{table*}

\begin{table*}[h!]
  \centering
  \caption{Experimental and theoretical magnetic-field positions, theoretical widths and relevant quantum numbers of the observed $^{53}$Cr$\ket{\text{a}}$-$^{53}$Cr$\ket{\text{b}}$ Feshbach resonances. The non-resonant scattering length remains constant at $a_{\rm bg}=39.6a_0$ to within \(0.5\%\). We assume an error on the experimental resonance position $B_{0,\text{exp}}$ of about 100 mG. All resonances have total angular momentum projection $M=-8$ with the exception of $^*$ which has $M=-6$, corresponding to incoming $d$-waves.}
  \label{tab:FRsCr53}
    \begin{ruledtabular}
\begin{tabular}{cccccccccccccccc}
$B_{0, \text{exp}}$(G) & $B_{0,\text{th}}$(G) & $\Delta_{\rm th}$(mG)
& $\ell$ & $m$ & $S$ & $M_S$ & $I$ & $M_I$
& $f_a$ & $f_b$ & $m_{fa}$ & $m_{fb}$ & $f$ & $m_f$ & $\delta \mu(\mu_{\rm B})$  \\
\hline
\noalign{\vskip 2mm}
8.85 & 8.756 & 1.0  & 4 & & 4 & & 2 & & & & & & 6 &  & 3.3\\
18.22 & 19.026 & 455 & 2 & &  & & & & 9/2 & 9/2 & & & 8 & &1.3\\
200.53 & 200.451 & 20 &  2 & -2 & 4 & & 2 & & & & & & 6 & -6 & 3.8\\
282.67 & 283.956 & $<0.1$ &  4 & -4 & & & & & & & & & & -4 & 4.0\\
345.98 & 344.943 & 3.1 & 4 & -2 & & & & & & & & & & -6 &  3.8\\
366.01 & 365.477 & 3.1 &  4 & -2 & & & & & & & & & & -6 & 3.8\\
367.12 & 367.247 & 0.3 &  2 & -2 & 3 & -3 & 3 & -3 & & & & & & -6 & 5.7\\
586.03 & 584.944 & 61 & 2 & -2 & & -4 & & -2 & & & & & & -6 & 3.8\\
607.72 & 606.780 & 130 &  2 & -2 & & & & & & & & & & -6 & 3.7 \\
749.45 & 749.037 & 183 &  & & & & & & 7/2 & 9/2 & -7/2 & -7/2 & & -7 & 1.6 \\
785.93 & 785.422 & 53 &  4 & -1 & & & & & & & & & & -7 & 1.8 \\
1175.70 & 1173.646 & 1.3 & 2 & -2 & & & & & & & & & & -6 &  2.0\\
1245.40 & 1246.044 & 705 & 2 & -1 & & & & & 7/2 & 9/2 & -7/2 & -7/2 & &-7 &1.9\\
1275.49 & 1275.801 & $<0.1$ & 6 &-2 & 5 & -3 & 3 & -3 & & & & & & -6 & 6.1\\
1276.09 & 1276.536 & 599 &  2 & -1  &  & -5 &  & -2 & & & & & & -7 & 1.9 \\
%1276.09 & 1277.140 & $<0.1$ & 6 & &  &  &  &  & & & & & &  &  \\
1417.07 & 1415.772 & $ < 0.1$  & 4 & -3 & 4 & -4 & 2 & -1 & & & & & & -5 & 4.0 \\
1448.04$^*$ &1448.430 &  & 2 & & 5 & -2&  3 & & & & & & & & 8.0 \\
1456.91 & 1457.255 & 3.0 & 4 & -2 & 4 & -4 & 2 & -2 & & & & & & -6 & 4.0\\
1544.21 & 1545.845 & 1.5 & 0 & 0 & 5 & & 3 & & & & & & 8 & -8 & 2.0\\
\end{tabular}
    \end{ruledtabular}
\raggedright
\vskip -12pt
\end{table*}

\clearpage
\newpage

\section{Accurate measurement of the magnetic-field width of a selected FR} \label{app:UhlOrn}

\begin{figure}[H]\centering%------------------------------------------
\includegraphics[width=\columnwidth]{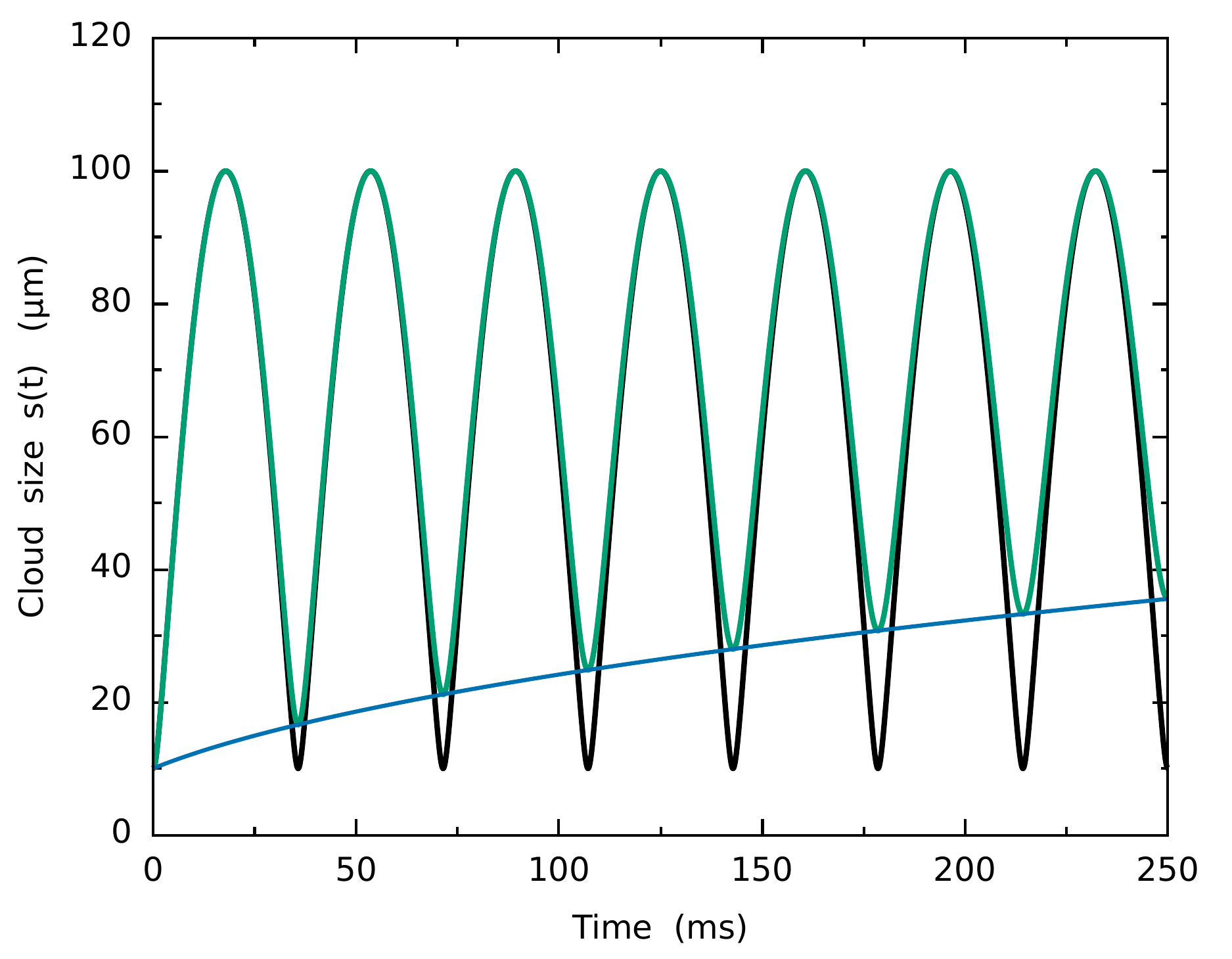}
\caption{Illustrative plot of Eq.~\eqref{eq:KetZwi} (black curve) and Eq.~\eqref{eq:UhlOrn} (green curve), together with the expected growth of the minimum size according to Eq.~\eqref{eq:UhlOrn_minsize} (blue curve).
The plot is obtained for $s_{0} = 10~\micron$, $\omega = 2\pi\times14~\Hz$, $\omega_0/\omega = s_{th}/s_{0} = 10$, and $\Gamma = 0.5~\secs^{-1}$, corresponding to $\Gamma/\omega \approx 0.0057$.}
\label{fig:UO_vs_KZ}
\end{figure}

In this appendix we present a novel technique to measure the magnetic-field width of a Feshbach resonance and, more generally, to characterize the elastic scattering rate around the zero-crossing of the scattering length.
The protocol is based on the high sensitivity of a specific type of breathing mode to weak damping factors, and it conveniently reduces the required number of measurements compared to standard methods \cite{RevModPhys.82.1225,PhysRevLett.129.093402,Guo2022,Krauser2017}.
The starting point is an ultracold gas confined in a harmonic trap.
If one trap frequency is instantaneously quenched to a lower value, a breathing mode of the cloud is triggered along the corresponding trap axis. 
In the absence of interactions, the resulting time evolution of the cloud size $\sx(t)$ is described by \cite{KetZwi2008,SykesBohn2015,Kagan1996}
\begin{equation}\label{eq:KetZwi}%--------------------------------------
\sx(t) =  \sx_0 \, \sqrt{\cos^2(\omega\:\! t) + \frac{\omega_0^2}{\omega^2}\,\sin^2(\omega\:\! t)} \,,
\end{equation}%---------------------------------------------------------
where $\sx_0=\sx(0)$, while $\omega_0$ and $\omega$ are the initial and final trap frequencies, respectively.
While numerical simulations are typically employed to take into account the effect of interactions \cite{SykesBohn2015,Wade2011,Sandro1999}, we realized that a more general analytic solution including collision-induced damping is offered by the seminal work of Uhlenbeck and Ornstein on the Brownian motion of a harmonically-bound particle \cite{UhlOrn1930}.
Their result reads
\begin{equation}\label{eq:UhlOrn}%--------------------------------------
\resizebox{0.85\linewidth}{!}{%
$\displaystyle \sx(t) = \sqrt{\sx_{th}^2 + (\sx_{0}^2-\sx_{th}^2)\,\ee^{-\Gamma t} \, \left(\cos(\Omega t) \,+\, \frac{\Gamma}{2\Omega}\, \sin(\Omega t)\right)^2} \,,$
}%
\end{equation}%---------------------------------------------------------
where $\sx_{th} = \sqrt{\frac{k_B T}{m \omega^2}}$ is the maximum size reached during the dynamics (corresponding to the asymptotic equilibrium value), $\Gamma$ is the collision-induced damping rate, and $\Omega = \sqrt{\omega^2 - \Gamma^2/4}$ is the effective oscillation frequency.
A peculiarity of Eq.~\eqref{eq:UhlOrn} lies in the fact that the maximum cloud size $\sx_{th}$ shows no damping, whereas the minimum one grows over time according to
\begin{equation}\label{eq:UhlOrn_minsize}%------------------------------
\sx_{\text{min}}(t) = \sx_{0}\, \sqrt{1\,+\,\left(\frac{\sx_{th}^2-\sx_{0}^2}{\sx_{0}^2}\right) \, \left(1 - \ee^{-\Gamma t} \right)}  \,.
\end{equation}%---------------------------------------------------------
In Fig.~\ref{fig:UO_vs_KZ} we show a comparison between Eqs.~\eqref{eq:KetZwi} (black curve) and \eqref{eq:UhlOrn} (green curve). One can notice that $\sx_{\text{min}}$ increases substantially after a few oscillation periods, even for relatively weak damping factors ($\Gamma/\omega\ll1$).

Given these insights, the idea behind our protocol is rather simple:
We excite a breathing mode in our ultracold sample, and we monitor the cloud size after an integer number of oscillation periods. 
By scanning the magnetic field close to a Feshbach resonance, we expect a minimum of $s(t=n\pi/\Omega \approx n\pi/\omega)$ at the zero-crossing of the scattering length, where the $s$-wave elastic scattering rate vanishes.

\begin{figure}[H]\centering%------------------------------------------
\includegraphics[width=\columnwidth]{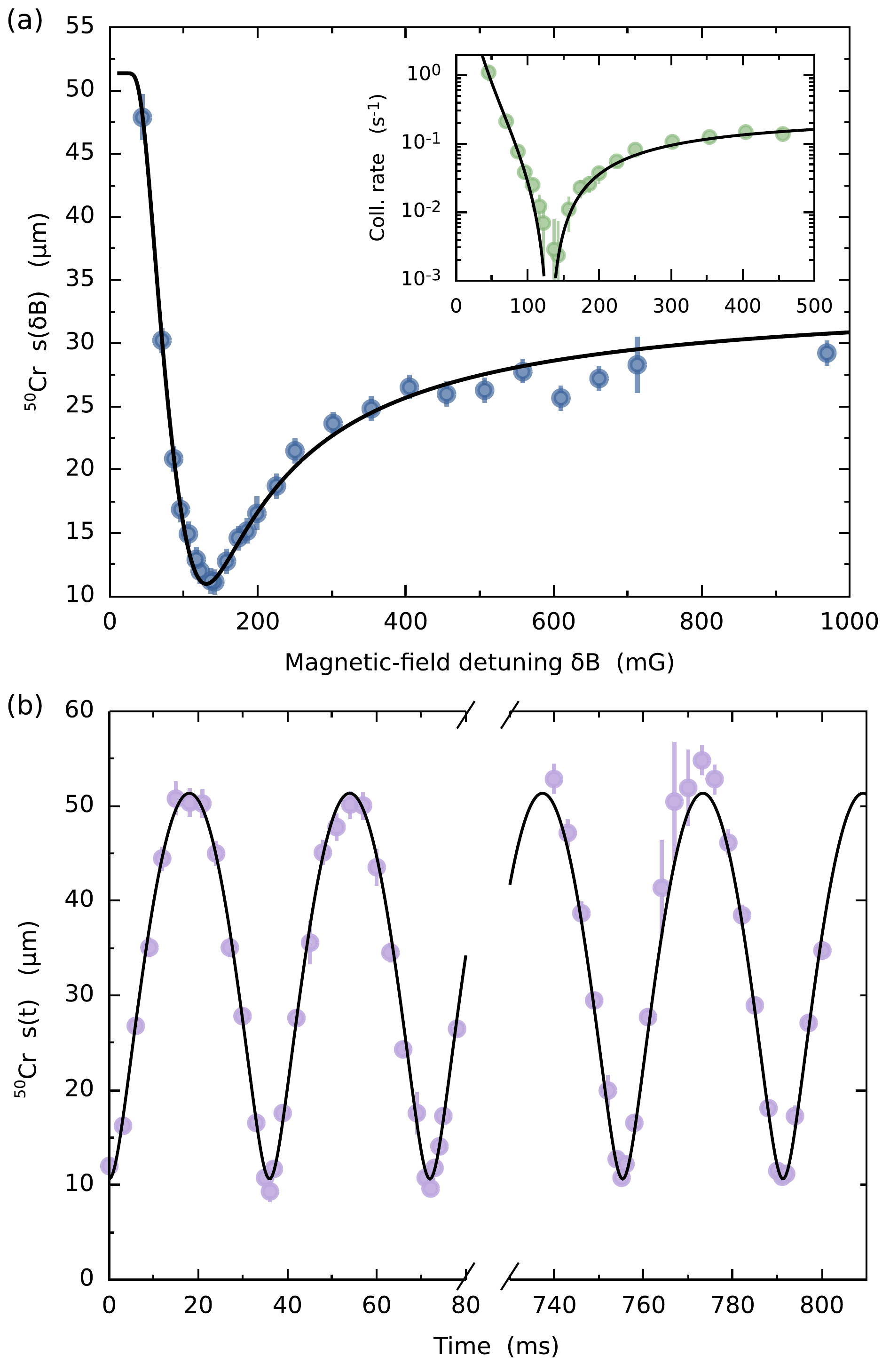}
\caption{ (a) Measured cloud size after $n=50$ oscillation periods (blue circles), as a function of the magnetic-field detuning $\delta B$. The black solid curve shows the best fit of Eq.~\eqref{eq:UhlOrn}, with only $\Delta_B$ and $\alpha$ as free parameters, see text for details.
Inset: Effective collision rate $\Gamma(\delta B)$ (green circles), as defined in Eq.~\eqref{eq:elscattrate}, extracted directly from the blue data according to Eq.~\eqref{eq:gammafromdata}. Here the black solid curve is a plot of Eq.~\eqref{eq:elscattrate} with the best-fitted values of $\Delta_B$ and $\alpha$.
(b) Undamped breathing oscillations recorded in proximity of the zero-crossing of the scattering length. The black solid line shows the best fit of Eq.~\eqref{eq:KetZwi}, from which we extract $\omega = 2\pi \times 13.902(2)$~Hz, $s_0 = 10.6(3)~\micron$, and $s_{th} = 51.4(3)~\micron$.}
\label{fig:UhlOrn_ZC_Cr50BEC}
\end{figure}

%---------------------------------------------------------
In the experiment, we prepare an ultracold gas of $^{50}$Cr$\ket{\rm a}$ ($T \approx 120$~nK), initially confined in the crossed ODT already described in the main text, at a magnetic-field detuning of about $+700$~mG above the FR located at $1390.7$~G.
The breathing mode is triggered by abruptly switching off the vertical beam, thereby allowing Cr to expand along the axis of the main ODT, similarly to what done for the $^6$Li-${}^{53}$Cr mixture in our Ref.~\cite{Finelli2026}. The magnetic-field curvature of the FESH1 coils provides nearly-ideal harmonic confinement, which is essential for preventing spurious damping caused by potential anharmonicity. 
The magnetic-field detuning $\delta B$ is adjusted with our FESH3 coils, whose contribution to the total curvature is negligible, and set to the target value $10$~ms before the start of the measurement.
We let the cloud evolve for about $1.8$~s (corresponding to $n=50$ oscillation periods) before performing \emph{in situ} absorption imaging to extract its size. 
As shown in Fig.~\ref{fig:UhlOrn_ZC_Cr50BEC}(a) (blue dots), we observe a clear trend of $s_{\text{min}}(\delta B)$, with a pronounced minimum around $\delta B\approx 125$~mG.
At this magnetic-field detuning, we recorded a time evolution of the breathing oscillations, which exhibited essentially no damping, as expected; see Fig.~\ref{fig:UhlOrn_ZC_Cr50BEC}(b).
We fitted Eq.~\eqref{eq:KetZwi} to these undamped oscillations to extract $s_0$, $s_{th}$, and $\omega$, which were then kept fixed in the following steps. 
The precise location of the resonance pole was determined from a loss measurement, similarly to those described in the main text.

The resonance magnetic width $\Delta_B$ can then be extracted by fitting Eq.~\eqref{eq:UhlOrn} to  Fig.~\ref{fig:UhlOrn_ZC_Cr50BEC}(a) data.
More specifically, we express the damping rate as
\begin{equation}\label{eq:elscattrate}%--------------------------------
\Gamma(\delta B) = \frac{\overline{n}\,\sigma(\delta B)\,v_{\text{rel}}}{\alpha}  \,,
\end{equation}%---------------------------------------------------------
where $\overline{n}$ is the density-weighted-density time-averaged over one period of Eq.~\eqref{eq:KetZwi}, $\sigma(\delta B) = 8\pi\, a^2(\delta B)$ is the elastic $s$-wave scattering cross section for indistinguishable bosons, with $a(\delta B) = a_{\text{bg}}(1-\Delta_B/\delta B)$, and $v_{\text{rel}} = \sqrt{\frac{16 k_B}{\pi} \frac{T}{m}}$ is the average relative velocity assuming thermal equilibrium \cite{Anderlini2005}. Finally, $\alpha$ is a dimensionless parameter, independent of $\delta B$, that both absorbs possible errors in the estimates of $\overline{n}$ and $v_{\text{rel}}$, and accounts for the effective number of collisions to damp the oscillations \cite{SykesBohn2015,Monroe1993}.
With these prescriptions, fixing %$a_{\text{bg}} = 47~a_0$ 
$a_{\text{bg}}$ to the theoretical value, see Table~\ref{tab:FRsCr50}, and leaving $\alpha$ and $\Delta_B$ as free parameters, we obtain the best fit shown in Fig.~\ref{fig:UhlOrn_ZC_Cr50BEC}(a) (black solid curve), which catches our experimental data fairly well.
The fit returns $\alpha=3.5(2)$ and $\Delta_B = 130(6)$~mG, where the former value appears reasonable compared to similar results found in the literature \cite{SykesBohn2015,Monroe1993}, and the latter agrees well with the theoretical prediction from our quantum collisional model (see Tab.~\ref{tab:FRsCr50}).
This shows that our collisional model captures the magnetic-field widths well, despite neglecting second-order spin-orbit interactions.

As a final remark, we emphasize that this experimental protocol can be equivalently exploited to extract the effective scattering rate directly from the data, without a fit, at least in the regime where $\Gamma\ll\omega$.
Indeed, from Eq.~\eqref{eq:UhlOrn}, setting $t=\thold=n\pi/\omega$ and $\Gamma\ll\omega$ (such that $\Omega\simeq\omega$), one finds
\begin{equation}\label{eq:gammafromdata}%-------------------------------
\Gamma(\delta B) = -\,\frac{1}{\thold}\,\log\left( \frac{s_{th}^2 - s(\delta B)}{s_{th}^2 - s_0^2}\right)  \,,
\end{equation}%---------------------------------------------------------
where $s(\delta B)$ is the measured cloud size at $t=\thold$.
We show an example of this procedure in the inset of Fig.~\ref{fig:UhlOrn_ZC_Cr50BEC}(a), where the (green) data points are obtained by converting $s(\delta B)$ according to Eq.~\eqref{eq:gammafromdata}, while the black solid line shows the Eq.~\eqref{eq:elscattrate} trend, with the best-fit parameters given above.
One can appreciate the sensitivity of this method to comparably weak scattering rates, on the order of (or even better than) $0.01~\secs^{-1}$. 

An analogous characterization has been performed for the $^{53}$Cr FR at 1245.4~G, yielding in this case $\Delta_B = 660(60)$~mG, again consistent with the theory prediction, see Table \ref{tab:FRsCr53}.

%---------------------------------------------------------

\end{document}